\documentclass[%
 reprint,
 superscriptaddress,
 nofootinbib,
 showkeys,
 amsmath,amssymb,
 aps,
 pra,
 floatfix,
]{revtex4-1}
\usepackage[T1]{fontenc}
\usepackage[utf8]{inputenc}
\usepackage{xargs}
\usepackage{graphicx}
\usepackage{booktabs,multirow,dcolumn,tabularx}
\usepackage{xspace,enumitem,fnpct}
    
\usepackage{amsmath,amssymb,amsthm}
\usepackage{mathtools,slashed,mathrsfs,esint}
\usepackage{hyperref}
\usepackage{cleveref}
\usepackage[automake,acronym]{glossaries-extra}
    \makeglossaries
    \glsdisablehyper
    \setabbreviationstyle[acronym]{long-short}
\usepackage{color,xcolor}
    \definecolor{wBlue}{rgb}{0.25,0.41,0.88}
    \definecolor{wGreen}{rgb}{0,0.5,0}
    \definecolor{wBrown}{rgb}{0.6,0,0}
    \definecolor{mBrown}{HTML}{604c38}
    \definecolor{mBlack}{HTML}{23373b}
    \definecolor{mOrange}{HTML}{EB811B}
    \definecolor{mGreen}{HTML}{14B03D}
    \definecolor{mBlue}{HTML}{4169E1}
    \definecolor{mRed}{HTML}{C91100}
    \definecolor{cBlue}{RGB}{137,188,188}
    \definecolor{cRed}{RGB}{205,137,137}
    \definecolor{cGreen}{RGB}{188,188,137}
    \definecolor{cGray}{RGB}{229,229,229}

\providecommand\name{}\renewcommand\name[1]{#1}

\providecommand\WSLDA{}\renewcommand{\WSLDA}
    {\href{https://wslda.fizyka.pw.edu.pl}
    {\emph{W-SLDA Toolkit}}\xspace}
\providecommand\ddcc{}\renewcommand{\ddcc}{\ensuremath{\lambda}}
\providecommand\kF{}\renewcommand{\kF}{\ensuremath{k_\mathrm{F}}}
\providecommand\eF{}\renewcommand{\eF}{\ensuremath{\varepsilon_\mathrm{F}}}

\providecommand\abinitio{}\renewcommand\abinitio{\emph{ab initio}\xspace}

\providecommand\aposteriori{}\renewcommand\aposteriori{\emph{a posteriori}\xspace}

\providecommand\etc{}\renewcommand\etc{etc.\xspace}
\providecommand\ie{}\renewcommand\ie{i.e.\xspace}
\providecommand\eg{}\renewcommand\eg{e.g.\xspace}
\providecommand\etal{}\renewcommand\etal{et al.\xspace}
\providecommand\cf{}\renewcommand\cf{cf.\xspace}

\providecommand\oprime{}\renewcommand\oprime{\ensuremath{^\prime}}
\providecommand\ooprime{}\renewcommand\ooprime{\ensuremath{^{\prime\prime}}}
\providecommand\odag{}\renewcommand\odag{\ensuremath{^\dag}}

\providecommand\ostar{}\renewcommand\ostar{\ensuremath{^\star}}

\providecommand\oast{}\renewcommand\oast{\ensuremath{^\ast}}

\providecommand\mo{}\renewcommand\mo{\ensuremath{^{-1}}}

\providecommand\I{}\renewcommand\I{\ensuremath{\mathrm{i}\mkern1mu}}
\providecommand\e{}\renewcommand\e{\ensuremath{\mathrm{e}\mkern1mu}}  

\let\qparen\relax\DeclarePairedDelimiter\parenq{\lparen}{\rparen}
\let\qbrack\relax\DeclarePairedDelimiter\brackq{\lbrack}{\rbrack}

\providecommand\qparen{}\renewcommand\qparen[1]{\parenq*{#1}}
\providecommand\qbrack{}\renewcommand\qbrack[1]{\brackq*{#1}}

\let\abs\relax\DeclarePairedDelimiter\abs{\lvert}{\rvert}

\let\function\relax
\DeclarePairedDelimiterXPP\function[2]{#1}{\lparen}{\rparen}{}{#2}
\let\functional\relax
\DeclarePairedDelimiterXPP\functional[2]{#1}{\lbrack}{\rbrack}{}{#2}

\providecommand\f{}\renewcommand\f[2]{\function*{#1}{#2}}

\let\pv\relax\DeclareMathOperator{\pv}{\mathcal{P}}
\let\order\relax
\DeclarePairedDelimiterXPP\order[1]{\mathcal{O}}{\lparen}{\rparen}{}{#1}
\providecommand\flatfrac{}\renewcommand\flatfrac[2]
{\ensuremath{\left.#1\middle\slash#2\right.}}
\providecommand\vb{}\renewcommand\vb[1]{\ensuremath{\boldsymbol{#1}}}
\providecommand\vbnabla{}\renewcommand\vbnabla{\ensuremath{\vb{\nabla}}}
\let\vbdot\relax\DeclareMathOperator{\vbdot}{\vb{\cdot}}

\let\grad\relax\DeclareMathOperator{\grad}{\vb{\nabla}}
\let\div\relax\DeclareMathOperator{\div}{\grad\vbdot}

\newcommand\vdsym{\ensuremath{\mathrm{d}}}
\newcommand\vDsym{\ensuremath{\mathcal{D}}}
\newcommand\vpsym{\ensuremath{\partial}}
\newcommand\vfsym{\ensuremath{\delta}}
\newcommand\vGsym{\ensuremath{\Delta}}

\newcommandx\vd[2][1={}]{\ensuremath{\mathop{\vdsym^{#1}#2}}}
\newcommandx\vD[2][1={}]{\ensuremath{\mathop{\vDsym^{#1}#2}}}
\newcommandx\vp[2][1={}]{\ensuremath{\mathop{\vpsym^{#1}#2}}}
\newcommandx\vf[2][1={}]{\ensuremath{\mathop{\vfsym^{#1}#2}}}
\newcommandx\vG[2][1={}]{\ensuremath{\mathop{\vGsym^{#1}#2}}}

\newcommandx\dvd[3][1={}]{\ensuremath{\frac{\vd[#1]{#2}}{\vd{#3}^{#1}}}}
\newcommandx\dvD[3][1={}]{\ensuremath{\frac{\vD[#1]{#2}}{\vD{#3}^{#1}}}}
\newcommandx\dvp[3][1={}]{\ensuremath{\frac{\vp[#1]{#2}}{\vp{#3}^{#1}}}}
\newcommandx\dvf[3][1={}]{\ensuremath{\frac{\vf[#1]{#2}}{\vf{#3}^{#1}}}}
\newcommandx\dvG[3][1={}]{\ensuremath{\frac{\vG[#1]{#2}}{\vG{#3}^{#1}}}}

\newcommandx\fdvd[3][1={}]{\ensuremath{\flatfrac{\vd[#1]{#2}}{\vd{#3}^{#1}}}}
\newcommandx\fdvD[3][1={}]{\ensuremath{\flatfrac{\vD[#1]{#2}}{\vD{#3}^{#1}}}}
\newcommandx\fdvp[3][1={}]{\ensuremath{\flatfrac{\vp[#1]{#2}}{\vp{#3}^{#1}}}}
\newcommandx\fdvf[3][1={}]{\ensuremath{\flatfrac{\vf[#1]{#2}}{\vf{#3}^{#1}}}}
\newcommandx\fdvG[3][1={}]{\ensuremath{\flatfrac{\vG[#1]{#2}}{\vG{#3}^{#1}}}}

\providecommand\op{}\renewcommand\op[1]{\ensuremath{\hat{#1}}}

\let\ket\relax\DeclarePairedDelimiter\ket{\lvert}{\rangle}
\let\braket\relax
\DeclarePairedDelimiterX\braket[2]{\langle}{\rangle}
    {#1 \delimsize\vert #2}
\let\ketbra\relax
\DeclarePairedDelimiterX\ketbra[2]{\vert}{\vert}
    {#1 \delimsize\rangle \delimsize\langle#2}
\let\expval\relax
\DeclarePairedDelimiterX\expval[2]{\langle}{\rangle}
    {#2 \delimsize\vert #1 \delimsize\vert #2}

\let\mel\relax
\DeclarePairedDelimiterX\mel[3]{\langle}{\rangle}
    {#1 \delimsize\vert #2 \delimsize\vert #3}
\let\commutator\relax
\DeclarePairedDelimiterX\commutator[2]{\lbrack}{\rbrack}{#1,#2}
\let\anticommutator\relax
\DeclarePairedDelimiterX\anticommutator[2]{\lbrace}{\rbrace}{#1,#2}
\let\innerproduct\relax
\DeclarePairedDelimiterX\innerproduct[2]{\langle}{\rangle}{#1,#2}
\let\normalordering\relax
\DeclarePairedDelimiterX\normalordering[1]{\vcentcolon}{\vcentcolon}{\mathop{}#1\mathop{}}

\let\timeordering\relax
\DeclarePairedDelimiterXPP\timeordering[1]{\mathcal{T}}{\lbrace}{\rbrace}{}{#1}

\makeatletter
\newcommand\DEFINEALPHABETLOOP[3]{%
  \ifx\relax#3\expandafter\@gobble
  \else\expandafter\@firstofone
  \fi
  {
  \expandafter\newcommand\expandafter*\csname#1#3\endcsname{#2{#3}}%
   \DEFINEALPHABETLOOP{#1}{#2}
  }
}
\newcommand\definealphabet[2]{%
  \DEFINEALPHABETLOOP{#1}{#2}abcdefghijklmnopqrstuvwxyzABCDEFGHIJKLMNOPQRSTUVWXYZ\relax
}%
\makeatother

\definealphabet{cal}{\mathcal}
\definealphabet{rm}{\mathrm}
\definealphabet{bb}{\mathbb}
\definealphabet{frak}{\mathfrak}
\definealphabet{scr}{\mathscr}
\definealphabet{vb}{\vb}

\newglossaryentry{gs}
{
    name={ground-state},
    description={ground-state}
}
\newglossaryentry{mb}
{
    name={many-body},
    description={many-body}
}
\newglossaryentry{qp}
{
    name={quasi-particle},
    description={quasi-particle}
}
\newglossaryentry{sp}
{
    name={single-particle},
    description={single-particle}
}
\newglossaryentry{UV}
{
    name={UV},
    description={ultra-violet}
}
\newacronym{ASLDA}{ASLDA}{Asymmetric Superfluid Local Density Approximation}
\newacronym{BCS}{BCS}{\name{Bardeen}\xspace– \name{Cooper}\xspace– \name{Schrieffer}}
\newacronym{BdG}{BdG}{\name{Bogoliubov}\xspace– \name{de Gennes}}
\newacronym{BEC}{BEC}{\name{Bose}\xspace– \name{Einstein} Condensate}
\newacronym{BHF}{BHF}{\name{Brueckner}\xspace– \name{Hartree}\xspace– \name{Fock}}
\newacronym{BMF}{BMF}{Beyond-Mean-Field}
\newacronym{DFT}{DFT}{Density Functional Theory}
\newacronym{DoF}{DoF}{Degrees of Freedom}
\newacronym{DR}{DR}{Dimensional Regularization}
\newacronym{EFT}{EFT}{Effective Field Theory}
\newacronym{EP}{EP}{Effective Potential}
\newacronym{EoS}{EoS}{Equation of State}
\newacronym{GF}{GF}{\name{Green} Function}
\newacronym{GGA}{GGA}{Generalized Gradient Approximation}
\newacronym{GLE}{GLE}{\name{Ginzburg}\xspace– \name{Landau} Equation}
\newacronym{GPE}{GPE}{\name{Gross}\xspace– \name{Pitaevskii} Equation}
\newacronym{HF}{HF}{\name{Hartee}\xspace– \name{Fock}}
\newacronym{HFB}{HFB}{\name{Hartree}\xspace– \name{Fock}\xspace– \name{Bogoliubov}}
\newacronym{HK}{HK}{\name{Hohenberg}\xspace– \name{Kohn}}
\newacronym{HST}{HST}{\name{Hubbard}\xspace– \name{Stratonovich} Transformation}
\newacronym{KS}{KS}{\name{Kohn}\xspace– \name{Sham}}
\newacronym{LEC}{LEC}{Low Energy Contsant}
\newacronym{LDA}{LDA}{Local Density Approximation}
\newacronym{LSE}{LSE}{\name{Lippmann}\xspace– \name{Schwinger} Equation}
\newacronym{MBGF}{MBGF}{Many-Body \name{Green} Function}
\newacronym{MBPT}{MBPT}{Many-Body Perturbation Theory}
\newacronym{MS}{MS}{Minimal Subtraction}
\newacronym{PDS}{PDS}{Power Divergence Subtraction}
\newacronym{QFT}{QFT}{Quantum Field Theory}
\newacronym{QMC}{QMC}{Quantum Monte Carlo}
\newacronym{RG}{RG}{Renormalization Group}
\newacronym{RPA}{RPA}{Random Phase Approximation}
\newacronym{SE}{SE}{\name{Schrödinger} Equation}
\newacronym{SCGF}{SCGF}{Self-Consistent \name{Green} Functions}
\newacronym{SFG}{SFG}{Superfluid \name{Fermi} Gas}
\newacronym{SLDA}{SLDA}{Superfluid Local Density Approximation}
\newacronym{TDBdG}{TDBdG}{time-dependent \name{Bogoliubov}\xspace– \name{de Gennes}}
\newacronym{TDHF}{TDHF}{time-dependent \name{Hartee}\xspace– \name{Fock}}
\newacronym{TDHFB}{TDHFB}{time-dependent \name{Hartee}\xspace– \name{Fock}\xspace– \name{Bogoliubov}}
\newacronym{UFG}{UFG}{Unitary \name{Fermi} Gas}
\newacronym{VPT}{VPT}{Variational Perturbation Theory}

\newcommand{\gw}[1]{\textcolor{red}{{\small GW: \textbf{ #1 } }}}

\usepackage[normalem]{ulem}

\newcommand{\orcidicon}[1]{\href{https://orcid.org/#1}{\includegraphics[height=\fontcharht\font`\B]{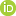}}}

\newcommand\update[1]{#1}

\newcommand\respadd[1]{#1}

\begin{document}
\preprint{}

\title{Local energy density functional for superfluid Fermi gases from effective field theory}

\author{Antoine Boulet\,\orcidicon{0000-0003-3839-6090 }}\email{antoine.boulet@pw.edu.pl}
\affiliation{Faculty of Physics, Warsaw University of Technology, Ulica Koszykowa 75, 00-662 Warsaw, Poland}

\author{Gabriel Wlaz\l{}owski\,\orcidicon{0000-0002-7726-5328}}\email{gabriel.wlazlowski@pw.edu.pl}
\affiliation{Faculty of Physics, Warsaw University of Technology, Ulica Koszykowa 75, 00-662 Warsaw, Poland}
\affiliation{Department of Physics, University of Washington, Seattle, Washington 98195--1560, USA}

\author{Piotr Magierski\,\orcidicon{0000-0001-8769-5017}}\email{piotr.magierski@pw.edu.pl}
\affiliation{Faculty of Physics, Warsaw University of Technology, Ulica Koszykowa 75, 00-662 Warsaw, Poland}
\affiliation{Department of Physics, University of Washington, Seattle, Washington 98195--1560, USA}

\date{\today}

\begin{abstract}
    Over the past two decades, many studies in the \acrlong{DFT} context revealed new aspects and properties of strongly correlated superfluid quantum systems in numerous configurations that can be simulated in experiments.
    This was made possible by the generalization of the \acrlong{LDA} to superfluid systems by \name{Bulgac} in [\href{https://doi.org/10.1103/PhysRevC.65.051305}{Phys. Rev. C \textbf{65}, 051305, (2002)}, \href{https://doi.org/10.1103/PhysRevA.76.040502}{Phys. Rev. A \textbf{76}, 040502, (2007)}].
    In the presented work, we propose an extension of the \acrlong{SLDA}, systematically improvable and applicable to a large range of many-body quantum problems getting rid of the fitting procedures of the functional parameters. 
    It turns out that only the knowledge of the density dependence of the \gls{qp} properties, namely,  the chemical potential, the effective mass, and the pairing gap function, are enough to obtain an explicit and accurate local functional of the densities without any adjustment \aposteriori. 
    This opens the way toward an \acrlong{EFT} formulation of the \acrlong{DFT} in the sense that we obtain a universal expansion of the functional parameters entering in the theory as a series in pairing gap function. 
    Finally, we discuss possible applications of the developed approach allowing precise analysis of experimental observations.
    In that context, we focus our applications on the static structure properties of superfluid vortices.
\end{abstract}

\maketitle

    The \gls{DFT} is a versatile method describing with very good accuracy the static, dynamic, and thermodynamic properties of \gls{mb} quantum systems in a unified framework \cite{Engel2011,Dreizler1990,Fiolhais2003,Koch2015,Messud2009,Parr1994,March1992,Engel2007}. 
    The success of this approach is due to its relatively low numerical cost compared to the methods that aim to solve under some well-controlled approximations the many-body Schr\"odinger equation. 
    The \gls{DFT} is one of the most popular methods in condensed matter physics, quantum chemistry, atomic physics, and nuclear physics due to its mathematical and conceptual simplicity.
    The essence of the modern \gls{DFT} relies on the \gls{KS} equations \cite{Kohn1965}, derived from the \gls{HK} theorem \cite{Hohenberg1964}, recasting the Schr\"odinger equation into a problem of non-interacting particles evolving in a density-dependent effective potential. 
    Although this procedure is exact, the form of the full effective potential remains unknown \cite{Becke2014}. 
    In the state of the art, the unknown (exchange-correlation) part of the functional is approximated \cite{Grasso2019,Bender2003,Saperstein2016} \eg using the so-called \gls{LDA} or the \gls{GGA} \cite{Luo2018,Ullrich1996,Burke1998,Burke1998a,Perdew1996,Becke1988,Lee1988,Becke1992a,Becke1992b,Becke1993}. 
    These procedures, guided by the \name{Landau} theory of \name{Fermi} liquid \cite{Landau1956,Landau1959,Baym2008,Lipparini2003} and its extension to finite systems by \name{Migdal} \cite{Migdal1967}, require the empirical or semi-empirical adjustment of the parameters appearing in the expansions, allowing an accurate description of the systems of interest.
    The \gls{KS} equations arise when densities are parametrized via {\it orbitals} $\phi_n$, which allow us to write the kinetic contribution in simple form $\sim\sum_n|\nabla \phi_n|^2$. In condensed matter, these types of functionals are typically referred as meta-GGA~\cite{PhysRevLett.91.146401}. Although, very accurate for describing systems being in the normal state, the standard KS approach is not able to deal with states exhibiting spontaneous symmetry breaking of the ground-state \cite{Ripka1986}.
    For instance, the breaking of the $U(1)$ symmetry, associated with the particle number conservation, allows to capture most of the \gls{BMF} static correlations such as superfluidity \cite{Anderson1972}.

    The challenging problem of the generalization of local \gls{DFT} to superfluid systems was achieved in \cite{Bulgac2007,Bulgac2012} for dilute (spin-symmetric and imbalanced) fermionic systems at unitarity, \ie when the $s$-wave scattering length of the bare two-body interaction becomes large.
    This development was guided by (i) the so-called \gls{BdG} theory or \gls{HFB} approximation \cite{Gennes1999,Martin2004,Cyrot1973,Leggett1980,Fetter2003} for weakly interacting systems in \gls{BCS} regime \cite{Bardeen1957} and (ii) the absence of other scales, except mean inter-particle distance, at unitarity \cite{Zwerger2012}. 
    A regularization scheme has been introduced and extensively discussed in \cite{Bulgac2002, Bulgac2002a, Yu2003} to remove \gls{UV} divergences of the pairing fields arising from the fact that local effective contact pairing interaction is considered to build the functional.
    This last point was an essential component to solve numerically the generalized \gls{BdG} equations, arising from the functional minimization.
    From there, numerous studies of the properties of quantum \name{Fermi} systems followed and revealed unexpected collective phenomena while providing a better understanding of experimental observations in ultracold atoms physics or nuclear physics \cite{Wlazlowski2018,Bulgac2014,Bulgac2013,Bulgac2016,Stetcu2011,Bulgac2019,Magierski2019,Georgescu2014,Bloch2008,Lewenstein2007,Bloch2012,Zwerger2012,Chin2010,Navon2010,Ku2012}.
   
  The purpose of this work is twofold.
  First, it consists of a general strategy of constructing local \gls{DFT} and getting rid of fitting procedures strongly depending on the system considered, \ie defining the functional parameters explicitly in terms of the physical quantities of the systems such as the \gls{qp} properties.
  Second, we 
  revisit the renormalization schemes of effective density-dependent (pairing) contact interaction usually used which do not take into account the \emph{in-medium} effects and the presence of the \name{Fermi} sea.
  Finally, the constructed functional is applied to describe static properties of superfluid quantum vortices which are crucial for understanding the dynamical processes observed in recent experiments \cite{Kwon2021}.


\section{Local \gls{DFT} for superfluid systems}\label{sec:local-DFT}

    We consider an unpolarized system of interacting fermions with equal masses $m$ in natural units ($m = \hbar  = k_\rmB = 1$) labeled by their spin-projection number $\sigma \in \{a,b\}$ on the quantization axis. 
    In the (local) \gls{DFT} formalism, the ground-state energy of the superfluid is given by
\begin{align}\label{eq:local-functional}
	E =& \int \calE(n(\vbr),\tau(\vbr),\nu(\vbr))\vd{\vbr}.
\end{align}
The (local) energy-density $\calE$ is a function of the normal ($n$), kinetic ($\tau$), and anomalous ($\nu$) densities constructed from the \name{Bogoliubov} \gls{qp} wave-functions expressed as a doublet \name{Nambu} spinor $\psi_n\odag = (u_n\oast,v_n\oast)$ that satisfy the generalized \gls{BdG} equations $\calH \psi_n = E_n \psi_n$ for positive eigenvalues.
    The effective density-dependent grand-canonical Hamiltonian reads
\begin{align} \label{eq:KS-effH}
	\calH & =                    
	\begin{bmatrix}
	    K + U - \mu & \Delta  \\
	    \Delta\oast     & -K\oast - U\oast  + \mu
	\end{bmatrix},
\end{align}
where $\mu$ denote the chemical potential.
    The densities are then given in terms of the \name{Bogoliubov} amplitudes as follows:
\begin{subequations} \label{eq:bdg-densities}
	\begin{align}
		n(\vbr)
		& = 2\sum_{E_n>0}\left(\abs{u_{n}(\vbr)}^2f^+_n + \abs{v_{n}(\vbr)}^2f^-_n\right), \label{eq:bdg-densities-n}
		\\
		\tau(\vbr)
		& = 2\sum_{E_n>0}\left(\abs{\grad u_{n}(\vbr)}^2f^+_n + \abs{\grad v_{n}(\vbr)}^2f^-_n\right), \label{eq:bdg-densities-tau}
		\\
		\nu(\vbr)
		& = \sum_{E_n>0} (f^-_n - f^+_n)u_{n}(\vbr)v_{n}\oast(\vbr) ,\label{eq:bdg-densities-nu}
	\end{align}
\end{subequations}
where the \name{Fermi} -- \name{Dirac} distribution is noted as $f_{n}^\pm = [1 + \exp \left ( \pm\beta E_n \right )]\mo$, 
\respadd{ with $\beta = 1/T$ being inverse of temperature. Here, we construct the functional for zero-temperature limit, however one can extend the formalism beyond this limit by introducing the thermal weights $f_{n}^\pm$ to the densities. Note that this extension of DFT concept to finite temperatures is approximate. It is equivalent to the finite-temperature \gls{HFB} theory~\cite{GOODMAN198130}. Also, finite and small temperature is frequently introduced to the numerical scheme, in order to improve converge properties of a self-consistent algorithms.}
	The kinetic, potential, and pairing operators are given by varying the functional with respect to the density fields, \ie $K(\vbr) = -\div \fdvf{E}{\tau(\vbr)} \grad $, $U(\vbr) = \fdvf{E}{n(\vbr)}$, and $\Delta\oast(\vbr) = - \fdvf{E}{\nu(\vbr)}$ respectively.

    We introduce a typical momentum scale of the many-body \name{Fermi} system that we identified as being the \name{Fermi} momentum $\kF$ related to the total density as $n \equiv \kF^3/3\pi^2$ and the associated \name{Fermi} energy $\eF \equiv \kF^2/2$. 
    Then, we assume that the energy-density is given by the following general form:
\begin{align} \label{eq:sldae-functional}
	\calE & = A_\ddcc \frac{\tau}{2}
	        + \frac{3}{5}B_\ddcc n \eF
	        + \frac{C_\ddcc }{n^{1/3}} \abs{\nu}^2,
\end{align}
where the \emph{functional parameters} $\{A_\ddcc,B_\ddcc,C_\ddcc\}$ are functions of the density-dependent coupling constant $\ddcc \sim \kF$.
The terms have the following physical meaning: the first one describes kinetic energy, the second one is related to the interaction energy, and the last one models the pairing correlations responsible for the superfluid properties of the system. 
\respadd{As discussed later, this local form of functional \eqref{eq:sldae-functional} leads to divergences of the kinetic (first term) and anomalous density (third term) independently that require proper renormalization scheme in order to cancel these UV divergences in the energy density expression. In practice, we introduce a cutoff energy to the sum appearing in eqs~(\ref{eq:bdg-densities}) and the coupling constant $C_\ddcc / n^{1/3}$ is renormalized according to the value of the cutoff.}
    In the following, we concentrate on dilute interacting systems for which only the leading order of the two-body interaction contributes and we set the dimensionless density-dependent coupling constant $\lambda = \abs{a_s\kF}$ where $a_s$ refers to the $s$-wave scattering length of the bare Hamiltonian. 
Note that, in general the coupling constant is defined through local value of Fermi momentum $\kF(\vbr)$, and thus in general all functional parameters are position dependent, through the coupling constant dependence $\lambda(\vbr)$. 

    In context of diluted \name{Fermi} gas with attractive two-body interaction ($a_s < 0$), we can mention that the so-called \gls{SLDA} and the \gls{BdG} functionals are efficient respectively close to unitarity ($\lambda \gg 1$), and in the weak coupling regimes ($\lambda \ll 1$)~\cite{Bulgac2012}.
Both of them can be described by the functional~(\ref{eq:sldae-functional}) upon proper choosing of the functional parameters, see also \cref{tab:notations}. 
    The main motivation of this work is to extend both of these functionals for a finite value of the $s$-wave scattering length, \ie obtain the correct density dependence of the functional parameters keeping the limiting regimes valid. Naively, in the simplest case, we can assume \name{Pad\'e} approximations of the functional parameters. Unfortunately, such approximations are inconsistent since the functional parameters are not independent, \ie linked to each other by physical constraints as the \name{Hugenholtz} -- \name{van-Hove} theorem \cite{Hugenholtz1958} for instance. To overcome this difficulty, we present below the strategy we adopt to obtain approximate expressions of the functional parameters in terms of the quasi-particle properties, \respadd{namely the effective mass, the chemical potential, and the pairing gap function.}
    However, it is important to notice that the proposed theory can be extended to a larger class of many-body systems with richer multi-body interactions. 

The popularity of the DFT method stems from the fact that it is a general-purpose method. It means that it can be applied to variety of setups, including  confined systems by an external potential $V_{\textrm{ext}}(\vbr)$ or coupled to the external pairing field $\Delta_{\textrm{ext}}(\vbr)$.
The functional~(\ref{eq:local-functional}) describes only intrinsic  energy of the system.  In the presence of external potentials, the total energy has the form 
\begin{align}
	E^\prime =& \int \calE(n(\vbr),\tau(\vbr),\nu(\vbr))\vd{\vbr} 
	+\int V_{\textrm{ext}}(\vbr)n(\vbr)\vd{\vbr}\nonumber\\
	&-\int\left(\Delta_{\textrm{ext}}(\vbr)\nu^*(\vbr)+\textrm{h.c.}\right)\vd{\vbr} .
\end{align}
Minimization of this functional introduces changes only to the mean-field and the paring potentials: $U(\vbr)\rightarrow U(\vbr)+V_{\textrm{ext}}(\vbr)$ and $\Delta(\vbr)\rightarrow \Delta(\vbr) + \Delta_{\textrm{ext}}(\vbr)$. Moreover in \cref{eq:local-functional}, we have assumed implicitly that the solution has no currents, $\vb{j}(\vbr)=0$, where
\begin{align}
\vb{j}(\vbr) = 2\sum_{E_n>0} \textrm{Im}[u^{*}_{n}(\vbr)\grad u_{n}(\vbr)]f^+_n \nonumber\\ 
-2\sum_{E_n>0} \textrm{Im}[v^{*}_{n}(\vbr)\grad v_{n}(\vbr)]f^-_n.
\end{align}
If the solution does not satisfy this requirement, for example, solution representing a quantum vortex, then one should use Galilean invariant expression for the kinetic density $\tau(\vbr)\rightarrow\tau(\vbr)-\vb{j}^{2}(\vbr)/n(\vbr)$, and add also to the total energy contribution from the matter flow
$E_{\textrm{flow}}=\int \frac{\vb{j}^{2}(\vbr)}{2n(\vbr)}\vd{\vbr}$, which accounts for center of mass motion energy~\cite{Bulgac2012}. In summary, in the presence of currents the functional \cref{eq:local-functional} must be redefined $E \to E + \Delta E$ where
\begin{align}
    \Delta E = \int (1 - A_\ddcc) \frac{\vb{j}^2(\vbr)}{2n(\vbr)} \vd{\vbr}.
\end{align}
    That induces a change in the definition of the kinetic and the potential operator appearing in the Hamiltonian of \cref{eq:KS-effH} as
\begin{subequations}
\begin{align}
    K(\vbr) &\to K(\vbr)
    - \frac{\I}{2} 
    \qbrack{
    {\dvf{\Delta E}{\vbj(\vbr)}}\vbdot{\grad}
    +
    {\grad} \vbdot {\dvf{\Delta E}{\vbj(\vbr)}}
    },
    \\
    U(\vbr) &\to U(\vbr)
    + \dvf{\Delta E}{n(\vbr)}.
\end{align}
\end{subequations}

\subsection{Constraints on quasi-particle properties}

    We first consider a homogeneous \name{Fermi} gas at zero-temperature of density $n$, ground-state energy per unit volume $E \equiv 3n\xi_\ddcc \eF/5$ and a chemical potential $\mu/\eF = \zeta_\ddcc$ that verify the thermodynamic relationship $\zeta_\ddcc = \xi_\ddcc + (\ddcc/5) \xi_\ddcc\oprime$.
    Moreover, we assume that the pairing gap function can be expressed as $\Delta \equiv - C_\ddcc \nu/n^{1/3} = \eta_\ddcc \eF$.
    Following the assumptions of the \gls{SLDA} approach relying on the \gls{BCS} theory results, we define the dispersion relation for the quasi-particle energies $E_k^2 = {\varepsilon_k^2  + \Delta^2}$ with the quadratic approximate \gls{sp} energies $\varepsilon_k = k^2/2m\ostar + U - \mu$ where $k$ denote the momentum of the quasi-particle considered.
    The effective mass $m\ostar$ and the effective mean-field potential $U$ are obtained by varying the energy functional according to the densities $\tau$ and $n$, respectively. 
    \respadd{In order to maintain compatibility with notation of the original SLDA functional~\cite{Bulgac2007}}, we define the \emph{\gls{SLDA} parameters} $\{\alpha_\ddcc,\beta_\ddcc,\gamma_\ddcc\}$ as $\flatfrac{1}{m\ostar} \equiv \alpha_\ddcc$ and $U \equiv (\tau/2 ) \fdvp{\alpha_\ddcc}{n}  + \beta_\ddcc \eF - \Delta^2/(3n^{2/3}\gamma_\ddcc)$ with
\begin{subequations} \label{eq:def-abc}
    \begin{align}
		\alpha_\ddcc
		& = A_\ddcc 
		\\
		\beta_\ddcc
		& = B_\ddcc + \frac{\ddcc}{5} \dvp{}{\ddcc}B_\ddcc,\label{eqn:beta-B}
		\\
		\frac{1}{\gamma_\ddcc}
		& = \frac{1}{C_\ddcc} + \ddcc \dvp{}{\ddcc}\frac{1}{C_\ddcc},
	\end{align}
\end{subequations}
where we have use chain rules of derivatives $3n\fdvp{X}{n} = \ddcc\fdvp{X}{\ddcc}$.
    Note that the \gls{SLDA} parameters are related to the functional parameters $B_\ddcc$, and $C_\ddcc$ defining the functional in \cref{eq:sldae-functional} but in general they differ from each other except
    for the unitary point, where $B_\infty = \beta_\infty$ and $C_\infty = \gamma_\infty$.
The \gls{sp} energies reads $\varepsilon_k = {\alpha_\ddcc k^2}/2 + b_\ddcc \eF$ with the shorthand notations
$b_\ddcc = (\tau/2 ) \fdvp{\alpha_\ddcc}{n} + (\beta_\ddcc - \zeta_\ddcc) - (3\pi^2)^{2/3}\eta_\ddcc^2/6\gamma_\ddcc$, and we introduce $c_\ddcc = 6C_\ddcc/(3\pi^2)^{2/3}$ that will be used below, defining the \emph{\gls{HFB} parameters}. \respadd{As it will be shown later, these parameters have compact representation in terms of quantities that are accessible for quantum Monte Carlo calculations, eqs~(\ref{eq:bc-expand}). Here, there are used to shorten the notation.}
    For the reader, we provide in \cref{tab:notations} a summary of the notations and conventions for the parameters used throughout the document. 
    
    The \gls{BCS} theory stands that the following integral equations are fulfilled \cite{Bulgac2007,Bulgac2012}:
\begin{subequations} \label{eq:BCS}
	\begin{align}
		n \,\equiv\, \frac{\kF^3}{3\pi^2}\, & = \frac{1}{2\pi^2}\int k^2\vd{k} \qparen{1-\frac{\varepsilon_k}{E_k}},
		\label{eq:BCS1}
		\\
		\frac{2}{\pi^2}\frac{\kF}{c_\ddcc}
		\equiv
		\frac{n^{1/3}}{C_\ddcc}
		& = \frac{1}{2\pi^2} \int k^2\vd{k} \qparen{\frac{1}{\alpha_\ddcc k^2} - \frac{1}{2E_k}}.
		\label{eq:BCS2}
	\end{align}
\end{subequations}
The meaning of these equations is following: the first one is zero temperature expression for the particle density, the \cref{eq:bdg-densities-n}, with explicit BCS formulas for $u_k$ and $v_k$, \respadd{while the second one corresponds to the gap equation where a counterterm was added to regularize integral as usually done in \gls{EFT} formulation of \gls{BCS}-like theory \cite{Papenbrock1999}.} We remark that \cref{eq:BCS1} is independent of $c_\ddcc$ and allows us to determine $b_\ddcc$ as a function of $\alpha_\ddcc$ for fixed value of $\eta_\ddcc$.
    Consequently, the effective mass parameter should be known to finally fix the value of $c_\ddcc$ with \cref{eq:BCS2}.
    From this fitting strategy, the result for $\eta_\infty = 0.493(12)$ corresponding to unitary regime has been obtained in \cite{Bulgac2007,Bulgac2012} with $\alpha_\infty = 1.094(17)$ and $\xi_\infty = 0.40(1)$.

\begin{table}
	\caption[Notations and conventions]{Summary of the notations and conventions used in this document. 
	The last column provides the equation(s) used to estimate the associated quantity/parameter with the proposed method. 
	The ground-state energy is given in unit of the Free Gas energy per unit volume $E_{\rmF\rmG} = 3n\eF /5$. 
	For reference, we give the value of the parameters at unitarity and at zero-density limit obtained using the APS$[x,y,z]$ functional used in this work.
	\label{tab:notations}}
	\centering
	\begin{tabularx}{\columnwidth}{@{\extracolsep{\fill}}clcrrc@{}}
		\toprule
		\multicolumn{2}{c}{APS$[x,y,z]$}                                     &
		                                                                     &
		\multicolumn{1}{c}{$\ddcc \to \infty$}                               &
		\multicolumn{1}{c}{$\ddcc \to 0$}                                    &
		\\ \midrule
		\multicolumn{1}{l}{ground-state energy}                              &
		\multicolumn{1}{l|}{$[E_{\rmF\rmG}]$}                                &
		$\xi_\ddcc$                                                          &
		$0.36$                                                               &
		\multicolumn{1}{r|}{$1$}                                             &
		\multirow{3}{*}{\eqref{eq:APSfunctional}} \\
		\multicolumn{1}{l}{chemical potential}                               &
		\multicolumn{1}{l|}{$[\eF]$}                            &
		$\zeta_\ddcc$                                                        &
		$0.36$                                                               &
		\multicolumn{1}{r|}{$1$}                                             &
		\\
		\multicolumn{1}{l}{effective mass}                                   &
		\multicolumn{1}{l|}{$[m]$}                                           &
		$m\ostar$                                                          &
		$1.19$                                                               &
		\multicolumn{1}{r|}{$1$}                                             &
		\\
		\multicolumn{1}{l}{pairing gap function}                             &
		\multicolumn{1}{l|}{$[\varepsilon_F]$}                               &
		$\eta_\ddcc$                                                         &
		$0.46$                                                               &
		\multicolumn{1}{r|}{$\frac{8}{\e^2}\exp\left(-\frac{\pi}{2\ddcc}\right)$}                                             &
		\eqref{eq:eta-parametrization} \\ \midrule
		\multicolumn{2}{c|}{\multirow{3}{*}{\textbf{functional parameters}}} &
		$A_\ddcc$                                                            &
		$0.84$                                                               &
		\multicolumn{1}{r|}{$1$}                                             &
		$= 1/m\ostar$ \\
		\multicolumn{2}{c|}{}                                                &
		$B_\ddcc$                                                            &
		$-0.28$                                                              &
		\multicolumn{1}{r|}{$-\frac{10}{9\pi}\ddcc$}                                             &
		\multirow{2}{*}{\eqref{eq:functional-functions}} \\
		\multicolumn{2}{c|}{}                                                &
		$C_\ddcc$                                                            &
		$-14.96$                                                             &
		\multicolumn{1}{r|}{$-\frac{4\pi}{(3\pi^2)^{1/3}}\ddcc$}                                             &
		\\ \midrule
		\multicolumn{2}{c|}{\multirow{3}{*}{\textbf{SLDA parameters}}}       &
		$\alpha_\ddcc$                                                       &
		$0.84$                                                               &
		\multicolumn{1}{r|}{$1$}                                             &
		$= 1/m\ostar$ \\
		\multicolumn{2}{c|}{}                                                &
		$\beta_\ddcc$                                                        &
		$-0.28$                                                              &
		\multicolumn{1}{r|}{$-\frac{4}{3\pi}\ddcc$}                                             &
		\multirow{2}{*}{(\ref{eq:def-abc})} \\
		\multicolumn{2}{c|}{}                                                &
		$\gamma_\ddcc$                                                       &
		$-14.96$                                                             &
		\multicolumn{1}{r|}{$\frac{15\pi^3/(1-7\ln2)}{(3\pi^2)^{1/3}\ddcc}$}                                             &
		\\ \midrule
		\multicolumn{2}{c|}{\multirow{2}{*}{\textbf{HFB parameters}}}        &
		$b_\ddcc$                                                            &
		$-0.62$                                                              &
		\multicolumn{1}{r|}{$-1$}                                            &
		\multirow{2}{*}{\eqref{eq:bc-expand}} \\
		\multicolumn{2}{c|}{}                                                &
		$c_\ddcc$                                                            &
		$-9.38$                                                              &
		\multicolumn{1}{r|}{$-\frac{8}{\pi}\ddcc$}                                             &
		\\ \bottomrule
	\end{tabularx}
\end{table}

    From \cite{Papenbrock1999,Furnstahl2007,Marini1998}, we can evaluate these integrals analytically in \gls{DR} with \gls{MS} scheme, \update{for instance}, by defining
\begin{align} \label{eq:Legendre-def}
	I_l(s,t) & \equiv -\fint_0^\infty \frac{z^l \vd{z}}{\sqrt{(z+s)^2 + t^2}}
	\nonumber\\
	& = \frac{\pi}{\sin \pi l} ({s^2+t^2})^{l/2} \f{P_l}{s/\sqrt{s^2+t^2}},
\end{align}
where $P_l$ denote the \name{Legendre} functions of the first kind and \update{the sign $\fint$ denote the integration in DR + MS scheme}.
    Note that in \gls{MS} scheme, integrations of powers of $z$ are assumed to give zero contribution to the final results, \ie $\fint z^l\vd{z} \to 0$. 
    After a change of variables and assuming $\alpha_\ddcc > 0$, \cref{eq:BCS} can be rewritten as:
\begin{subequations}
	\label{eq:BCS-Legendre}
	\begin{align}
		\label{eq:BCS-Legendre1}
		1
		& =
		\frac{3}{4}\qbrack{\f{I_{3/2}}{\frac{b_\ddcc}{\alpha_\ddcc},\frac{\eta_\ddcc}{\alpha_\ddcc}}+ \qparen{\frac{b_\ddcc}{\alpha_\ddcc}} \f{I_{1/2}}{\frac{b_\ddcc}{\alpha_\ddcc},\frac{\eta_\ddcc}{\alpha_\ddcc}}},
		\\
		\label{eq:BCS-Legendre2}
		\frac{\alpha_\ddcc}{c_\ddcc}
		& = 
		\frac{1}{8} \f{I_{1/2}}{\frac{b_\ddcc}{\alpha_\ddcc},\frac{\eta_\ddcc}{\alpha_\ddcc}}.
	\end{align}
\end{subequations}
    \update{Remark that, these analytic results are independent from the regularization scheme used to calculate the integrals of \cref{eq:BCS}.}
    We solve \cref{eq:BCS-Legendre} perturbatively by expanding in power of $(\eta_\ddcc/\alpha_\ddcc)^n$ as follows\footnote{For stability reason, due to the fact that we simply use quadratic approximation for the \gls{sp} energies, we consider only the case $b_\ddcc<0$ \cite{Volovik2007,Pankratov2012,Kaiser2013}.}:
\begin{subequations} \label{eq:bc-expand}
	\begin{align}
		b_\ddcc
		& = \alpha_\ddcc \sum_n \calB_n \qparen{\ln \eta_\ddcc/\alpha_\ddcc} \times \qparen{\frac{\eta_\ddcc}{\alpha_\ddcc}}^{n},
		\\
		\frac{1}{c_\ddcc}
		& = \frac{1}{\alpha_\ddcc} \sum_n \calC_n \qparen{\ln \eta_\ddcc/\alpha_\ddcc} \times \qparen{\frac{\eta_\ddcc}{\alpha_\ddcc}}^{n},
	\end{align}
\end{subequations}
where the first functions $\calB_n$ and $\calC_n$, up to $n = 8$, are given in \cref{app:Legendre}. 
We obtained prescription for inducing HFB parameters from the effective mass $m\ostar=1/\alpha_\ddcc$ and, the pairing gap $\Delta=\eta_\ddcc\eF$. The chemical potential $\zeta_{\lambda}$, and related to it the equation of state $\xi_{\lambda}$, as well as the kinetic density are needed in order to disentangle \gls{SLDA} parameters from HFB parameters. 
More precisely, the kinetic density is obtained in the same way as the normal density \cref{eq:BCS1} 
leading to:
\begin{align*}
    \frac{\tau}{n\varepsilon_\rmF} &= \frac{3}{2} 
    \qbrack{\f{I_{5/2}}{\frac{b_\ddcc}{\alpha_\ddcc},\frac{\eta_\ddcc}{\alpha_\ddcc}}+ \qparen{\frac{b_\ddcc}{\alpha_\ddcc}} \f{I_{3/2}}{\frac{b_\ddcc}{\alpha_\ddcc},\frac{\eta_\ddcc}{\alpha_\ddcc}}}.
\end{align*}
In practice, the $I_l(s,t)$ functions are approximated following the \cref{app:Legendre}.
    The expansion defined by \cref{eq:bc-expand} allows us to get rid of fitting procedures since the \gls{HFB} parameters are given by a systematically improvable series. 
    
The \cref{eq:bc-expand} together with definition of   \gls{SLDA} parameters provides prescription for inducing $\{\alpha_\ddcc,\beta_\ddcc\,\gamma_\ddcc\}$ from $\{m\ostar,\zeta_\ddcc,\eta_\ddcc\}$. In the last step we need to convert the \gls{SLDA} parameters into the functional parameters $\{A_\ddcc,B_\ddcc\,C_\ddcc\}$. 
\respadd{While for $A_\ddcc$ and $C_\ddcc$ it is trivial, for $B_\ddcc$ we need to solve differential equation~(\ref{eqn:beta-B}).
Its general solution}
can be expressed, after successive integrations by parts, as
\begin{align*}
	B_\ddcc
	& = \frac{5}{\ddcc^5}
	\int_{0}^\ddcc \beta_{l} \, {l}^{4} \vd l
	=
	\sum_{n = 0}^\infty (-1)^n \frac{5! \ddcc^n}{(n+5)!} \dvp[n]{\beta_{\ddcc}}{\ddcc}.
\end{align*}
Here we will consider the expansion only up to second order in $\ddcc \sim \kF$, \ie
\begin{align} \label{eq:functional-functions}
	B_\ddcc \simeq \beta_{\ddcc} - \frac{\ddcc \beta_{\ddcc}\oprime}{6}
	+ \frac{\ddcc^2 \beta_{\ddcc}\ooprime}{42}.
\end{align}
    As it will be shown, the truncated expansion is consistent with the second-order constraint on the weak coupling regime, and provides numerically accurate results.

    Summarizing, in order to extract the $\ddcc$ dependence of the functional parameters, we need:
\begin{enumerate}
	\item the total ground-state energy, or equivalently the chemical potential, \ie the generalized \name{Bertsch} parameter, for all values of $\ddcc$, \eg those developed in \cite{Lacroix2016,Adhikari2008};
	\item the $\ddcc$-dependence of the associated effective mass, \eg discussed in \cite{Boulet2019} related to \abinitio adjustments;
	\item the evolution of the pairing gap function as a function of $\ddcc$ at zero-temperature obtained form \abinitio calculations and experiments.
\end{enumerate}
    Consequently, we argue that the proposed method is applicable not only for diluted systems but also for general superfluid \gls{mb} systems, providing we have access
    (from either experiments or \abinitio calculations)
    to the chemical potential, the effective mass, and the pairing field without further adjustment.
    Strictly speaking, the parametrization of the induced functional is unique and universal, assuming that the functional has SLDA-type form, \cref{eq:sldae-functional}. Although, the functional has been constructed to reproduce properties of the uniform system, it is expected to provide good quality results also for non-uniform systems, as it was in the case of the original SLDA functional~\cite{Bulgac2012}.  
    In the next section, we describe the methodology to impose the correct limits and obtain these physical quantities.

\subsection{Parameterization of physical quantities for homogeneous dilute systems at zero-temperature \label{subsec:APSxyz}}

\begin{figure*}\centering
	\includegraphics{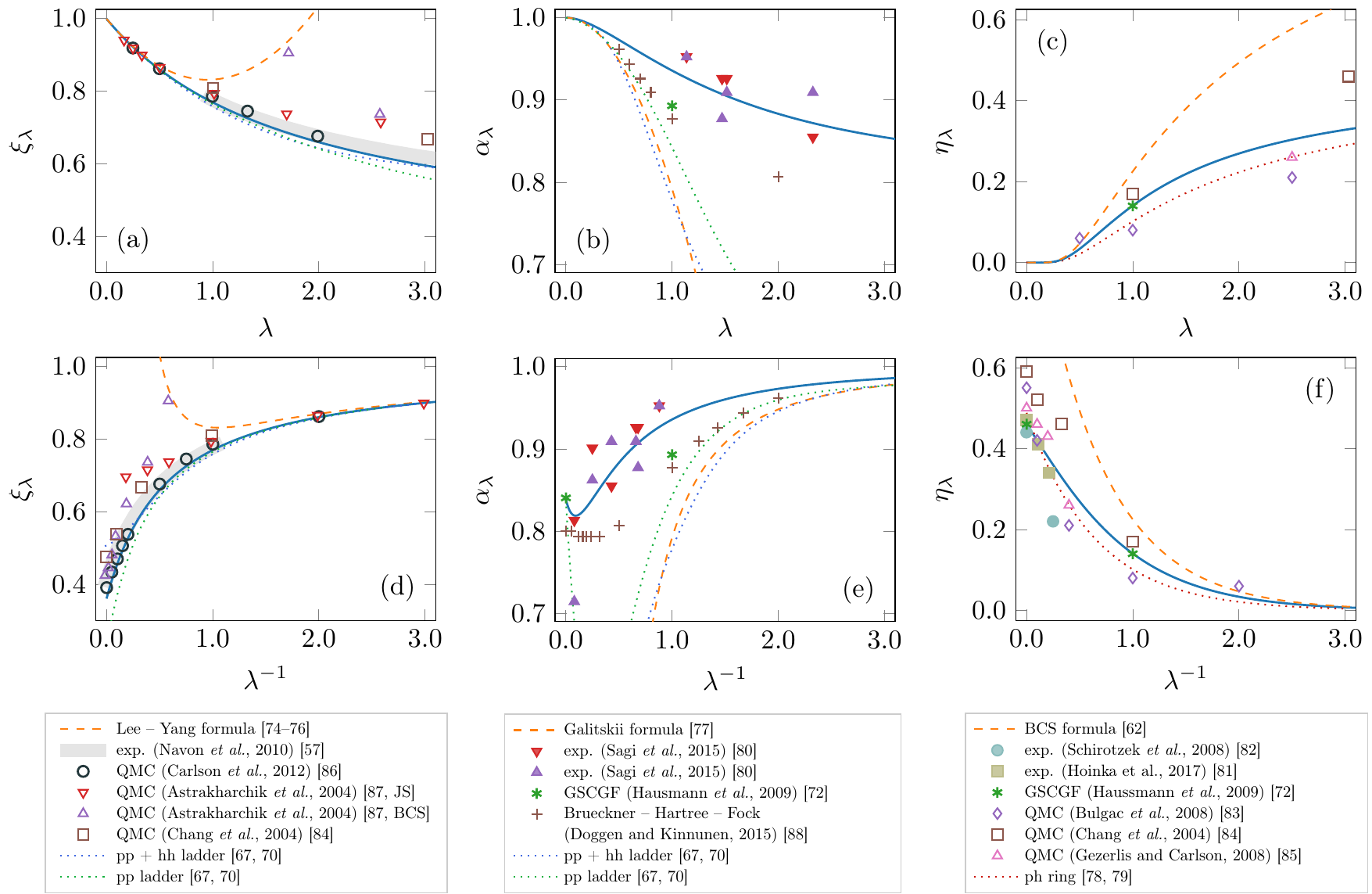}
\caption[SLDAE parameters]
	{
	Dimentionless ground-state energy [first column (a, d)], effective mass [second column (b, e)], and pairing gap function [third column (c, f)] for spin-symmetric infinite dilute systems of identical fermions with negative $s$-wave scattering length at zero-temperature as a function of $\ddcc$ [first row (a, b, c)] and $\ddcc\mo$ [second row (d, e, f)].
	The blue solid line correspond to our parametrization of the functional defined by \cref{eq:sldae-functional}: the ground-state energy and the effective mass correspond to the APS functional \cite{Boulet2019} by setting $x \simeq -0.75$ adjusted to reproduce the result of \cite[green asterisks]{Haussmann2009} at unitarity, and the paring gap function is given by \cref{eq:eta-parametrization}.
	For reference, we display the second order \gls{MBPT} results, known as the \name{Lee} -- \name{Yang} \cite{Huang1957,Huang1957a,Lee1957} and \name{Galitskii} formula \cite{Galitskii1958} for the ground-state energy and the effective mass respectively, and the \gls{BCS} equation \cite{Papenbrock1999} (orange dashed lines) as well as the ladder resummation of the particle-particle \cite{Boulet2019,Kaiser2013} (green dotted lines), particle-particle and hole-hole \cite{Boulet2019,Kaiser2013} (blue dotted lines), and particle-hole channels \cite{Gorkov1961,Chen2016} (red dotted line). 
	We show also various experimental results from \cite[triangles]{Sagi2015}, \cite[light green squares]{Hoinka2017}, \cite[light blue circles]{Schirotzek2008}, 
	\cite[grey area]{Navon2010}, \gls{QMC} calculations \cite[purpule open diamonds]{Bulgac2008}, \cite[brown open squares]{Chang2004}, \cite[pink open up-triangles]{Gezerlis2008},
	\cite[black open circles]{Carlson2012}, \cite[red open down-triangles (\name{Jastrow} -- \name{Slater} trial wave function) and purple open up-triangles (\gls{BCS} trial wave function)]{Astrakharchik2004}, and
	\acrlong{BHF} calculation \cite[plus crosses]{Doggen2015}.
	}
	\label{fig:sldae-fit}
\end{figure*}

    For the ground-state energy or the generalized \name{Bertsch} parameter $\xi_\ddcc$, we chose the APS functional~\cite{Boulet2019} displayed in \cref{fig:sldae-fit}(a, d) and the associated inverse effective mass $\alpha_\ddcc$ (shown in \cref{fig:sldae-fit}(b, e)) is obtained through the derivative at the \name{Fermi} surface of the associated \gls{sp} potential.
    To provide its explicit form, let us first introduce the following parametric function \cite{Boulet2019a}:
\begin{align}
	\mathcal{S}_\ddcc & = (\sigma - 1) \arctan \qparen{\frac{\ddcc u}{1 + \ddcc v}},
\end{align}
where the spin degeneracy is denoted by $\sigma$ ($\sigma=2$ for spin symmetric systems).
    The dimensionless ground-state energy, the chemical potential, and the effective mass are then given in terms of this parametric function and its first derivative as follows:
\begin{subequations}
    \label{eq:APSfunctional}
	\begin{align}
		\xi_\ddcc
		= 1 &- \frac{16}{3\pi}\calS_\ddcc,      
		\\
		\zeta_\ddcc
		= 1 &- \frac{16}{3\pi} \calS_\ddcc 
		      - \frac{16\ddcc}{15\pi} \calS_\ddcc\oprime ,
		\\
		\alpha_\ddcc
		= 1 &+ {\frac{5uw- 7x u v }{9\pi u^2}}
		        \ddcc^2 \calS_\ddcc\oprime
		\nonumber\\
		      &+ {\frac{2w+7x v}{9\pi u^2}}
		        (1+\ddcc v) \ddcc^2 {\calS_\ddcc\oprime}^2,
	\end{align}
\end{subequations}
where the constants are given by $u = 5/24$, $v = 6(11-2\ln 2)/35\pi$, $w = 24(1-7\ln 2)/35\pi$. The free parameter of the effective mass parametrisation, $x$, is adjusted to reproduce the result of \cite{Haussmann2009} at unitarity, $\alpha_\infty\mo = 1.19$, \respadd {leading to $x \simeq -0.75$.}

    The advantages of these parametrizations, discussed extensively in \cite{Boulet2019a}, reside in the fact that they provide a proper reproduction of thermodynamical properties of dilute \name{Fermi} gas at zero-temperature across the whole range $a_s <0$ \cite{Boulet2018}.
    In particular, the low-density expansions up to second order in the \gls{MBPT}, for the energy as well as the effective mass, are fulfilled while keeping finite values at unitarity under a compact and explicit form in terms of the density-dependent coupling constant $\ddcc$.
    Note in particular that the \name{Bertsch} parameter predicted by the functional\footnote{The exact value reads:
\[\xi_\infty = 1 - \frac{16}{3\pi}\f\arctan{\frac{175\pi/144}{11-2\ln{2}}} \simeq 0.358.\]} ($\xi_\infty \simeq 0.36$) is consistent with the \name{Gorkov} \acrlong{GF} results of \cite{Haussmann2009}. 
    The associated effective mass is displayed in \cref{fig:sldae-fit}(b, e) and compared to experimental data and theoretical results.
    
    The pairing gap function parametrization is then assumed to be consistent with the result of \cite{Haussmann2009} obtained at unitarity ($\eta_\infty = 0.46$) as well as the \gls{BCS} theory results. In the limit $\ddcc \ll 1$ (BCS regime) we have $\eta_\ddcc \sim ({8}/{\e^2}) \exp(-{\pi}/{2\ddcc}) \equiv \bar{\eta}_\ddcc$ \cite{Papenbrock1999,Furnstahl2007,Marini1998}.
    Therefore, we use the following parametrization for the paring gap function:
\begin{align} \label{eq:eta-parametrization}
	\eta_\ddcc
	= \frac{8}{\e^2} \f\exp{-\frac{\pi}{2\ddcc}}
	\times
	\frac{1+\ddcc y}{1+  \ddcc y z},
\end{align}
where $z \equiv {\bar{\eta}_\infty}/{\eta_\infty}$ and the free parameter $y$ is set arbitrarily to $y = 4/5$ in order to reproduce \abinitio calculations, see \cref{fig:sldae-fit}(c, f).
    Note that this parametrization of the pairing gap function under a \name{Pad\'e} approximation is \emph{empirical} (but suggested by the parametrization of the ground-state energy) and, as far as we know, remains to be validated from the ground.
    The parametrization of the ground-state energy, the effective mass and the pairing gap function as defined above will be denoted by APS$[x,y,z]$ and we chose the values $x \simeq -0.75$, $y = 4/5$, and $z = ({8}/{\e^2})/0.46$ in this work.

\begin{figure}
	\includegraphics{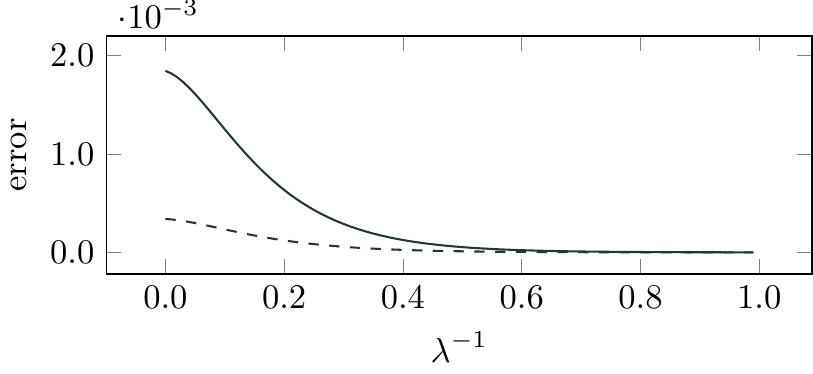}
	\caption[Numerical errors on \gls{BCS} equations]{
	Numerical error using perturbative solutions given by \cref{eq:bc-expand} in \cref{eq:BCS-Legendre} as a function of $\ddcc\mo$.
	The solid and the dashed lines show the absolute difference of the RHS and LHS of \cref{eq:BCS-Legendre1} and \cref{eq:BCS-Legendre2} respectively.
	Note that the perturbative expansion becomes less accurate, but still very reasonable, close to the unitary regime because $\eta_\infty/\alpha_\infty \sim 0.6$.}
	\label{fig:bc-err}
\end{figure}

    We have obtained the $\ddcc$-dependence of the physical quantities: the ground-state energy $\xi_\ddcc$ or equivalently the chemical potential $\zeta_\ddcc$, the associated effective mass $\alpha_\ddcc$, and the pairing gap function $\eta_\ddcc$.
    Thus, it allows us to determine 
    the \gls{HFB} parameters $b_\ddcc$ and $c_\ddcc$ using the perturbative approximation given by \cref{eq:bc-expand}. In \cref{fig:bc-err} we display relative error between results provided by the expansion \eqref{eq:bc-expand} and exact results of \cref{eq:BCS-Legendre}. 
    We observe that close to the unitary regime, the quality of the approximation provided by \cref{eq:bc-expand} 
    decreases. 
    This is due to the fact that, close to the unitary limit, the perturbative parameter $\eta_\ddcc/\alpha_\ddcc \simeq 0.6$ becomes large. 
    However, we can add more terms in the expansions to improve systematically the results. From practical point of view, extraction of the HFB parameters with the relative error below $1\%$ is sufficient. 
    
\begin{figure}
	\includegraphics{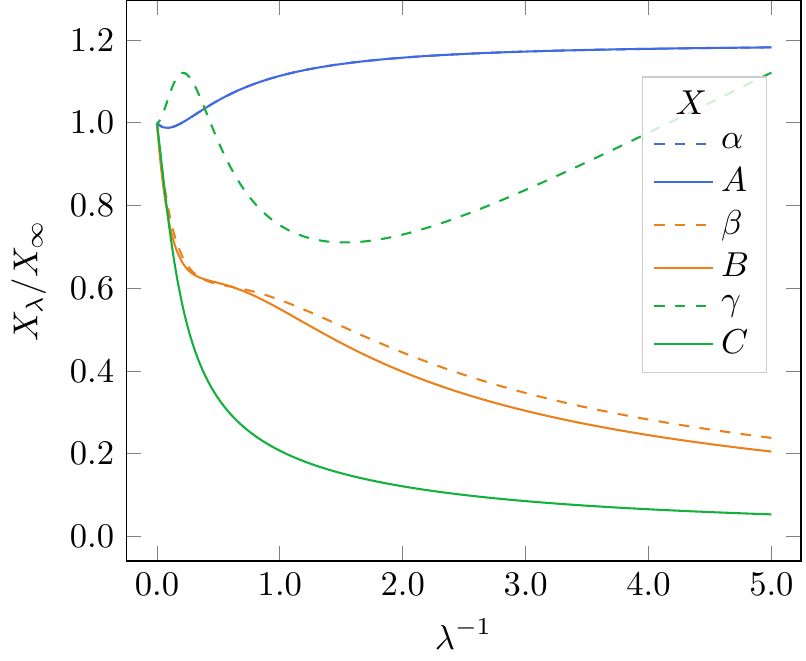}
	\caption[Functional parameters]
	{
	Normalized functional parameters $A_\ddcc$ [blue (upper) solid line], $B_\ddcc$ [orange (middle) solid line], and $C_\ddcc$ [green (lower) solid line] obtained with \cref{eq:functional-functions} and the approximation of \cref{eq:bc-expand} as a function of $\ddcc\mo$.
	Note that the exact results obtained by solving \cref{eq:def-abc} numerically provide indistinguishable results.
	The zero order approximations $A_\ddcc \sim \alpha_\ddcc$, $B_\ddcc \sim \beta_\ddcc$, and $C_\ddcc\mo \sim \gamma_\ddcc\mo$ are also displayed by the dashed lines.
	}
	\label{fig:ABC-trucated}
\end{figure}

    The results for the functional parameters obtained from \cref{eq:functional-functions} and the expansions given by \cref{eq:bc-expand} are shown in \cref{fig:ABC-trucated}.
    Note that, numerically, we do not observe significant differences 
    due to the fact that we apply (i) the truncated expansion given by \cref{eq:functional-functions} or (ii) the full integration of the solution, using (a) the perturbative expansion of \cref{eq:bc-expand} or (b) the exact numerical results.  
    {For reference, we provide in \cref{app:improved-BdG} the weak limit coupling of the functional up to second order in $\lambda$.}

    The 
    systematically improvable expansion given by \cref{eq:bc-expand} allows overcoming fitting procedure of numerical results obtained by solving \cref{eq:BCS} or \cref{eq:BCS-Legendre}.
    Actually, the results are strongly dependent on the ground-state energy, the effective mass, and the pairing gap function chosen to describe the system.
    In other words, this work depends only on the precision 
    with which the physical quantities 
    can be extracted from experiment and/or theoretical calculations.
    Moreover, in this work, we discuss only the ultracold atomic gas, but the developed theory, as suggested by the quality of the \gls{DFT} approach to describe many-body systems, can be extended to nuclear physics, condensed matter, quantum chemistry, \etc
    For the systems of interest, the proposed method requires only the knowledge of the \gls{qp} properties as a function of the density of the associated homogeneous infinite systems at zero-temperature.

\subsection{Toward an \acrshort{EFT} formulation of the \gls{DFT}?}

    The expansion of the functional parameters 
    in terms of microscopic quantities (and their derivatives 
    with respect to the density), can be considered as the first step toward an \gls{EFT} formulation \cite{Weinberg1979} of the \gls{DFT} \cite{Furnstahl2012,Furnstahl2020,Grasso2016,Furnstahl2007,Hammer2000}.
    Actually, starting from a microscopic point of view, the knowledge of the \gls{qp} properties of the system can be used to obtain a universal systematic expansion of the functional parameters, that is to say, the coupling constant of the effective Hamiltonian. 
    In other words, starting from a \gls{sp} picture, the coupling constant for the low-energy degrees of freedom, \ie the classical density fields, have been obtained in a systematic improvable expansion 
    and explicitly expressed through \gls{qp} properties.

    Essentially, the local \gls{DFT} discussed here  treats the energy-density at first order by neglecting cross-coupling between the classical fields of the theory, \eg terms in $ \tau \cdot \nu $, $ \nu \cdot n $, \etc at next-to-leading order. 
    It could be interesting in  
    future to consider such terms by using the standard perturbation techniques of \gls{EFT} (\gls{mb} diagrams, \name{Feynman} rules, power counting, \etc) and/or introducing other collective fields allowing 
    for 
    spontaneous symmetry breaking of the ground states. 
    Besides, we can enrich the resulting functional by considering higher order terms beyond 
    the quadratic approximation of the \gls{sp} energies assumed in this work.
    
    The \gls{EFT} 
    formulation that we have 
    presented is valid for uniform systems for which the \gls{DR} associated to \gls{MS} is 
    sufficient to remove the UV divergences from the theory.
    However, while the 
    \update{regulated analytic form}
    is very convenient, it does not allow to grasp simply the subtleties of the renormalization procedure.
    Besides, for finite systems, such regularization leads to numerical instabilities.
    In the next section, we discuss more precisely renormalizations in the case of non-uniform and finite systems. That will provide a proper identification of the low- and high-energy scales of the \gls{EFT} expansion and
    will shed light 
    on the formal aspects of the presented theory.



\section{Regularization of pairing fields \label{sec:regularization}}
At the formal level, densities $\tau$ and $\nu$ as defined by \cref{eq:bdg-densities-tau} and (\ref{eq:bdg-densities-nu}) are divergent. For example, according BCS theory we have $u_k v^*_k=\frac{\Delta}{2\sqrt{\varepsilon_k^2 + \Delta^2}}$, and then
\begin{align} 
\nu = \int \frac{\vd{\vb{k}}}{(2\pi)^3}u_k v^*_k
 = \frac{1}{2\pi^2}\int_{0}^{\infty} \frac{k^2 \Delta \vd{k}}{2\sqrt{\varepsilon_k^2 + \Delta^2}}\rightarrow \infty,
\end{align}
since $\varepsilon_k\sim k^2/2$. However, the energy of the system must be finite, which means that in the combination
$A_{\lambda}\frac{\tau}{2}+\frac{C_{\lambda}}{n^{1/3}}|\nu|^2$
divergences cancel out (note that $C_{\lambda}<0$). Thus, the theory must be supplemented with prescription how to deal with the  divergences. 
    
    The regularization of the \gls{BdG} equations relies on the link between the coupling constant $C_\ddcc/n^{1/3} \to g$ appearing in \cref{eq:sldae-functional} and the vacuum two-body $s$-wave scattering length $a_s$ through the spherical integral renormalization scheme
\begin{align} \label{eq:BdG-regularization}
	\frac{1}{4\pi a_s} = \frac{1}{g} + \frac{1}{4\pi^2}
	\pv{\int \frac{k^2 \vd{k} }{e_k }},
\end{align}
where $\pv$ denote the \name{Cauchy} principal value, $e_k = k^2 /2$ are the un-shifted \gls{sp} energies.
    Note that formally, it is not the regularization of the \gls{BdG} equations but the regularization of contact interaction, \ie even in \gls{MBPT} for dilute \name{Fermi} system, this regularization is required due to the \gls{UV} divergence of the scattering particle-particle loop amplitude in the vacuum, see also \cref{app:EFT-reg}.
    Thus, this regularization is only valid in the vacuum 
    due to the standard scattering theory starting from the bare Lagrangian. 
    The difficulty arises from the fact that we start from an effective density-dependent or \gls{HFB} Lagrangian. Consequently, this renormalization scheme becomes obsolete and the regularization must be performed \emph{in-medium}.
    The first attempt, and as far as we know, the only approach used in that context, was developed by \name{Bulgac} \etal \cite{Bulgac2007, Bulgac2002, Bulgac2002a, Yu2003}. 
    Let us briefly present the strategy.
Guided by \cref{eq:BdG-regularization}, it was suggested to replace the coupling constant by a regularized one in order to define the pairing gap function $g\mo \rightarrow n^{1/3}/C_\ddcc^\mathrm{reg.}$, and to make the replacement: 
$(4\pi a_s)\mo \rightarrow n^{1/3}/C_\ddcc$ using the density coupling constants 
defined above.
    Consequently, replacing
    the free \gls{sp} energies $e_k$ of \cref{eq:BdG-regularization} by the density-dependent \gls{sp} energies $e_k \to \varepsilon_k$, we obtain the following renormalization scheme [see \cref{app:EFT-reg} for explicit derivation in \gls{EFT} picture]:
\begin{align} \label{eq:slda-regularization}
	\frac{n^{1/3}}{C_\ddcc}
	= \frac{n^{1/3}}{C_\ddcc^\mathrm{reg.}}
	+ 
	\frac{1}{4\pi^2} \pv\int \frac{k^2 \vd{k}}{\alpha_\ddcc k^2/2 + b_\ddcc\eF }
\end{align}
where 
$C_\ddcc^\mathrm{reg.}$ is the regularized density dependent coupling constant used in numerical calculations.
The integral is then computed using spherical cutoff $\int \rightarrow \int_{0}^{k_c}$. 
Then, the \gls{BdG} densities must be computed with the consistent cutoff, \ie the summations in \cref{eq:bdg-densities} are performed over the \gls{sp} states with eigenvalues $E_n < E_c\approx k_c^2/2$.
    Consequently, the functional parameter $C_\ddcc$ obtained in the previous section within \gls{DR} + \gls{MS} procedure is used during the regularization process only. Namely, the renormalized coupling constants are used when we solve the numerical \gls{BdG} equations that define in particular the pairing gap function as $\Delta = -\frac{C_\ddcc^\mathrm{reg.}}{n^{1/3}}\nu_c$ where subscript $c$ indicates that the density is calculated  with the energy cutoff, i.e. $\sum_{E_n>0}\rightarrow \sum_{E_c>E_n>0}$.

\subsection{Results and comparisons}
\begin{figure*}\centering
	\includegraphics{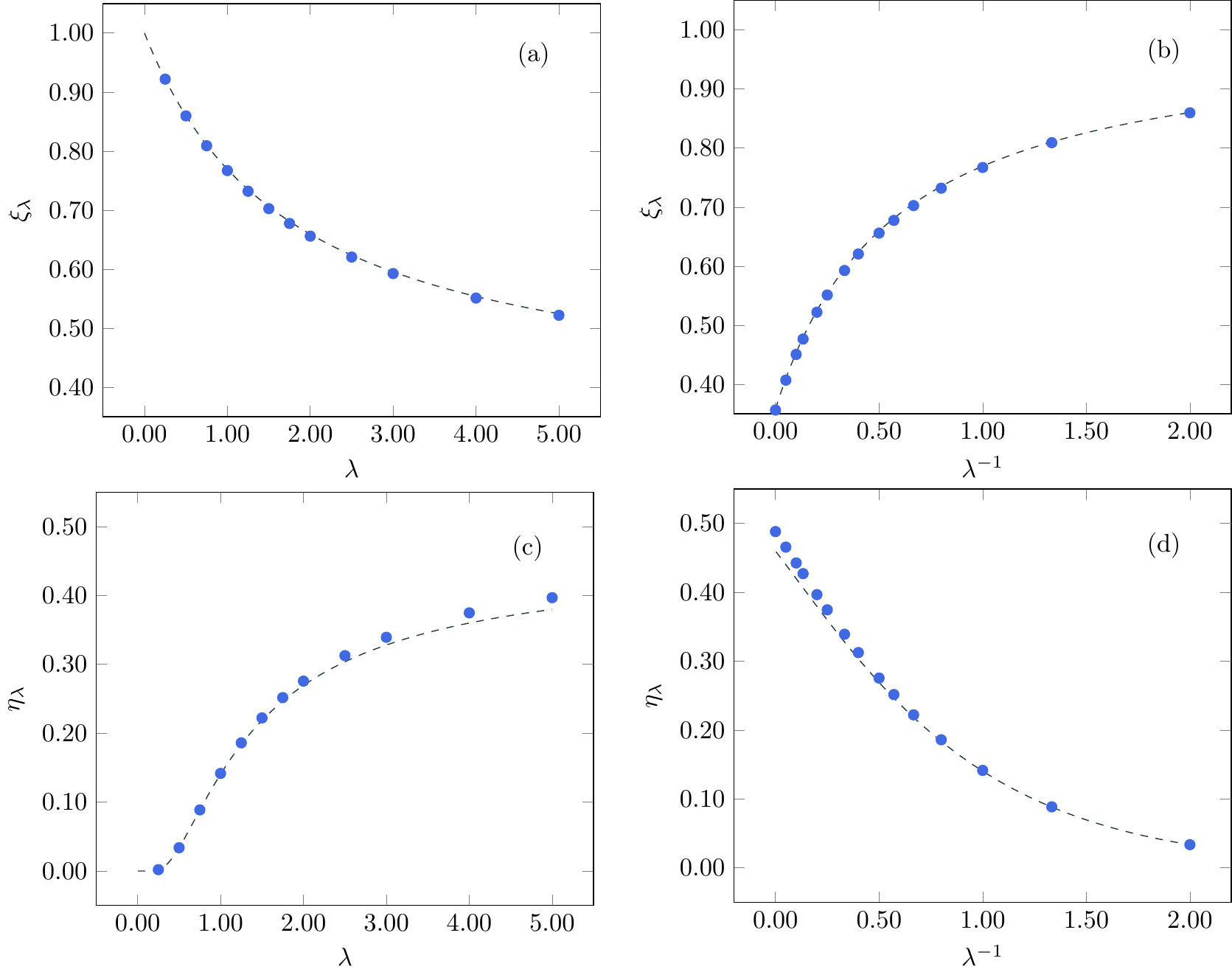}
	\caption[Ground state energy and pairing gap function]
	{
	Dimensionless ground-state energy [resp. pairing gap function] as a function of $\ddcc$ (a) [resp. (c)] and of $\ddcc^{-1}$ (b) [resp. (d)] obtained by solving numerically the \gls{BdG} equations associated to the constructed functional together with the regularization scheme defined by \cref{eq:slda-regularization} [blue filled circle]. \respadd{The computation was executed on a spatial mesh of size $80^3$ with lattice spacing $dx=1$, and density was set to satisfy $\kF=1$.}
	The dashed black line corresponds to the analytical parametrization of the APS$[x,y,z]$ functional that was used as the input for the induction procedure.
	}
	\label{fig:hxieta}
\end{figure*}

The constructed functional has been implemented within \WSLDA \cite{Wlazlowski2018,Bulgac2014}. The implementation is released as open-source via web page~\cite{WSLDAToolkit}.
    We have solved numerically the \gls{BdG} equations using the methods proposed in this work for spin symmetric homogeneous systems at zero-temperature for several values of the density-dependent coupling constant $\ddcc$. \respadd{The solver executes computation in 3-dimensional space, on a spatial grid of size $N^3$, with lattice spacing $dx$, which introduces natural energy cut-off scale $E_c=\frac{\pi^2}{2mdx^2}$.}
    The resulting ground-state energy and pairing gap function are shown in \cref{fig:hxieta}  
[blue filled circle]. 
    Note that numerical solution provided by the derived functional and supplemented with the regularization procedure reveals results in satisfactory agreement with the initial APS parametrization.
    The discrepancies of the pairing gap function (and the chemical potential) close to the unitarity  
    with respect to the analytic results that we used for inducing the functional parameters [black dashed line] can be seen. \respadd{They reflect systematic errors introduced by the computation process in discretaized space as well as by truncations applied in formulas~(\ref{eq:bc-expand}) and (\ref{eq:functional-functions}). The comparison provides a stringent test for the presented method: the functional parameters are generated by analytical formulas derived for continuum space and infinite cut-off energy, while the opposite process finds the self-consistent solution of eqs~(\ref{eq:KS-effH}) for discrete space and for finite energy cut-off.}
    Since the ground-state energy is well reproduced within the whole range from weak coupling to unitarity limit [see \cref{fig:hxieta}(a-b)], the thermodynamic properties of the \name{Fermi} gas at zero-temperature as well as the universal \name{Tan}'s contact parameter \cite{Zwerger2012,Tan2008,Tan2008a,Tan2008b} are consequently in good agreement with experimental observations (see \cite{Boulet2018} for more detailed discussions).

\begin{figure}\centering
	\includegraphics{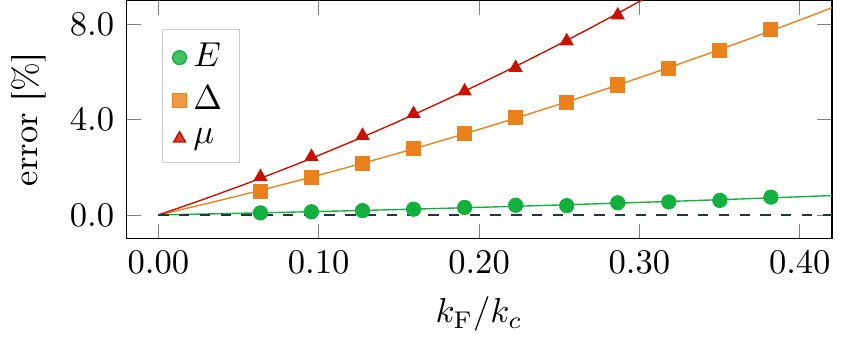}
	\caption[Error from cuttoff]
	{
 	Relative numerical error of the ground state energy [green filled circle], the pairing gap function [orange filled square], and the chemical potential [red filled triangle] as a function of $\kF/k_c$ obtained by solving the \gls{BdG} equations with the \WSLDA and constructed here functional at 
 	unitarity. 
	}
	\label{fig:error-kc}
\end{figure}

\subsection{Identification of scales and discussion}
    \respadd{Starting 
    from \cref{eq:bdg-densities}, we can show that, for large $k_c \gg k_\mathrm{F}$, $\tau \sim (\Delta / A_\ddcc)^2 k_c / \pi^2$ and $\nu \sim (\Delta / A_\ddcc) k_c / (2\pi^2)$.
    The renormalized pairing coupling constant is consistent with these results since $C_\lambda/n^{1/3} \sim -2\pi^2 A_\lambda /k_c$.} In particular, using the definition of the pairing gap function $\Delta = - C_\ddcc / n^{1/3} \nu$, we have
\begin{align}
    \frac{C_\ddcc}{A_\ddcc n^{1/3}} \abs{\nu}^2  = 
    -\frac{\Delta}{\eF\ostar} \nu^\ast \eF,
\end{align}
where we recognize the expansion parameter $x_\ddcc \equiv \eta_\ddcc / A_\ddcc = \flatfrac{\Delta}{\eF\ostar}$ with $\eF\ostar \equiv A_\ddcc \eF$ used in \cref{eq:bc-expand}. Thus, when looking at the problem from \gls{EFT} perspective we can identify $\Delta$ as the low-energy scale and $\eF\ostar$  as the high-energy scale.
Consequently, invoking only local pairing field in the description is justified when $\flatfrac{\Delta}{\eF\ostar}$ is small. 
This result is reminiscent of the pioneering work of Furnstahl, Hammer and Puglia \cite{Furnstahl2007} who used $\Delta/\eF$ as an expansion parameter in their \gls{EFT} for weak/dilute interacting systems. 
    Here, we have generalized 
    this approach by including \gls{qp} properties.
    {From this \gls{EFT} perspective, the leading order of the theory is equivalent to  the \gls{BdG} functional}
\begin{subequations}
\begin{align}\label{eq:BdG-functional}
    \calE_\text{BdG} = A_\ddcc \qparen{\frac{\tau}{2} -  x_\ddcc \nu\oast \eF} + \order{x_\lambda^2}.
\end{align}
    However, in the strong limit coupling, the pairing gap function is finite so that $x_\ddcc = \order{1}$. 
    In the SLDA functional the higher order corrections are modeled by terms related to $B_\ddcc$ parameter
\begin{align}
    \calE_\text{SLDA} =  A_\infty \qparen{\frac{\tau}{2} -  x_\infty \nu\oast \eF} + \frac{3}{5}B_\infty n \eF,
\end{align}
which implicitly depends on higher powers of $x$, \ie $B_\infty \sim \order{x_\infty^k}$.  
It is clear that $B_{\ddcc}$ is the only remaining parameter which should be expressed as a series in $x_\ddcc$ to have a proper \gls{EFT} formulation. To  
fulfill this requirement, we make use of the chain rule of derivatives, $\fdvd{}{\ddcc} = \fdvd{x_\ddcc}{\ddcc} \times \fdvd{}{x_\ddcc}$,
in \cref{eq:functional-functions}. Note also that the term proportional to $B_\ddcc$ quantifies difference between the \gls{BdG} functional (mean field) and the full functional. Thus, it may be treated as an analog of the so-called exchange-correlation term, widely discussed in standard \gls{DFT}. In general the $B_\ddcc$ does not vanish in the limit $x_\ddcc\rightarrow 0$ (transition to normal state), and then it models interaction effects that are not related to the pairing.

    Once the divergencies have been removed one may still wonder if all the cutoff dependencies are eliminated from the theory. Actually, in a proper \gls{EFT}, the energy density, and consequently the observables, must be independent of the cutoff momentum $k_c$ introduced to regularize the pairing coupling constant. 
    We start by fixing the cutoff momentum and then we evalute the density $n$ with \cref{eq:BCS1} such that we can introduce the \name{Fermi} momentum $\kF$, the density-dependent coupling constant $\ddcc$, and the associated \name{Fermi} energy $\eF$. 
    Subsequently, we define the energy density of the associated non-interacting system $\calE_0 = 3n\eF/5$, and using the \gls{BCS} results,
we can show that
\begin{align}
    A_{\ddcc} \qparen{\frac{\tau_c}{2} -  x_{\ddcc} \nu_c\oast \eF} 
    = \qbrack{\xi_{\ddcc} - B_{\ddcc}}\calE_0,
\end{align}
\end{subequations}
    where $\tau_c$ and $\nu_c$ are the cutoff-dependent renormalized densities. Finally, we can conclude that the divergences have been properly canceled.
{In \cref{fig:error-kc} we display numerically obtained the energy, the pairing gap and the chemical potential dependence on the cutoff parameter $k_c$. We find that indeed the total energy does not exhibit significant cutoff dependence. However, other two observables admit residual cutoff dependence. This residual cutoff dependence is also present in original \gls{SLDA} functional and its origin requires further investigation in future. We emphasize that the 
discrepancy between obtained and expected result vanish at zero density ($\kF \to 0$), which is formally equivalent to the infinite cutoff limit.}    

    To conclude, the formal development presented in this article allows us to go further in the future studies by using \gls{EFT} perturbative techniques, \ie \gls{mb} \name{Feynman} diagrams associated with the identification of a power counting, to enrich the functional by including higher-order terms in the energy-density, \eg at the second order, we expect terms as $(\Delta\oast\nu)^2$. 
    Moreover, further development could include higher orders in the gradient expansion and the effective range effects \cite{Forbes2012,Gezerlis2008,Lacroix2016,Lacroix2017,Boulet2018,Schonenberg2017,Schwenk2005,Kaiser2012}, \ie beyond quadratic approximation of \gls{sp} energies, generalize the \gls{BCS} \gls{qp} dispersion relation, consider spin-imbalance systems, \etc
    Despite the recent and impressive theoretical developments in \gls{SCGF} techniques \cite{Loos2018,Tarantino2017,Phillips2014,Soma2011,Haussmann2009,Haussmann2006,VanHoucke2012,Rossi2018,Rossi2018a} or \gls{QMC} approach of the many-body problems \cite{Bulgac2008,Chang2004,Gezerlis2008,Carlson2005,Astrakharchik2004} as well as experimental works \cite{Sagi2015,Hoinka2017,Schirotzek2008,Biss2021,Weimer2015,Horikoshi2017},
    precise constraints on the \gls{qp} properties are still limited. We still lack a general framework that allow 
    for systematically improvable 
    form of such quantities starting from the bare interaction. Among recent attempts 
    to obtain a \gls{DFT} from first principle, one can refer to the \gls{EFT} for dilute \name{Fermi} gas \cite{Schaefer2005,Hammer2000,Furnstahl2007,Furnstahl2008,Furnstahl2000,Steele2000,Steele2000a,Platter2003}, the bilocal Legendre transforms techniques \cite{Cornwall1974,Furnstahl2012,Drut2010,Polonyi2002}, the \gls{DFT} driven by \abinitio calculations \cite{Salvioni2020,Boulet2019} which did not lead to satisfying results \cite{You2000,You2003}.
    However, the \gls{SLDA} have proven
    to provide the formidable precision in the description of strongly correlated Fermi gases, despite its astonishing simplicity. Nevertheless, the use of the standard \gls{SLDA} so far was limited to the unitary regime which makes it sometimes 
    difficult to compare directly with the experiments. As we will discuss in the next section, the \gls{DFT} proposed in this work has the potential to reconcile the theoretical simulations and experimental results in ultracold atomic physics.
    


\section{Applications}

    In the previous sections, 
    we have introduced a methodology to construct systematically a \gls{SLDA}-like functional from 
    the density-dependent quasi-particle properties (the chemical potential, the effective mass, and the pairing gap function) only. 
We have focused 
on diluted \name{Fermi} systems
for which we have (i) introduced the APS$[x,y,z]$ parametrization of the functional in \cref{subsec:APSxyz}, (ii) truncated the expansions defined by \cref{eq:bc-expand} up to $n=8$, (iii) applied the approximations of \cref{eq:functional-functions}, and (iv) used the regularization scheme for the pairing field in \cref{sec:regularization}. Altogether, this \gls{DFT} will be called SLDAE for \emph{\gls{SLDA} Extended}. 
To illustrate the possibilities that are offered by
our approach, 
we 
provide some applications of the SLDAE functional. The implementation of SLDAE functional is publicly accessible via \WSLDA~\cite{WSLDAToolkit}. 
In the Supplemental Material~\cite{SupplementlMaterial} we provide detailed information about the computation process.

\subsection{Phase diagram and critical temperature}
\begin{figure*}
\centering
\includegraphics{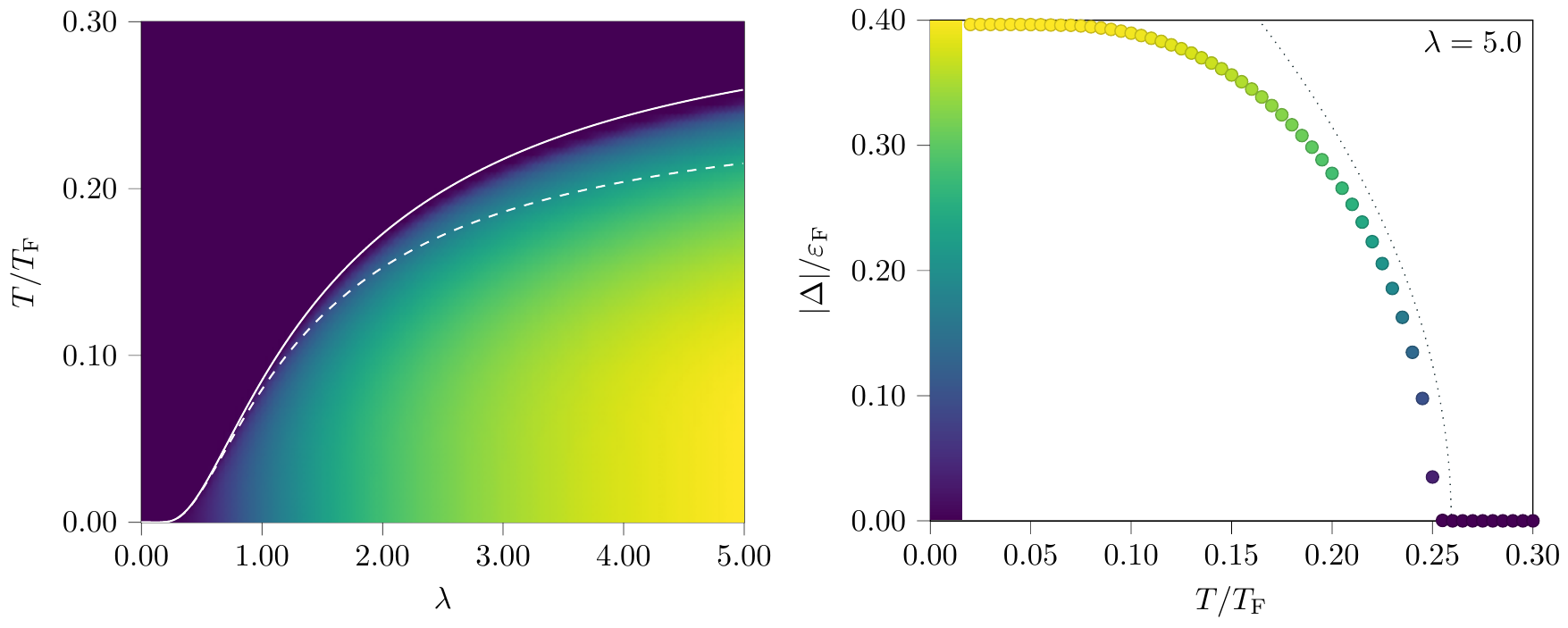}
\caption{
Phase diagram obtained using the SLDAE functional. On the left: projection of the dimensionless pairing gap function on the $(T/T_\rmF,\ddcc)$ plane. 
The dashed white line corresponds to the \gls{BCS} critical temperature formula valid in the weak coupling regime, \respadd{$T_c^{\textrm{BCS}} = (\mathrm{e}^\gamma/\pi) \Delta(T=0)$}.
We observe that the numerically extracted critical temperature is in very good agreement with \cref{eq:TcSLDAE} [white solid line].
On the right: dimensionless pairing gap function at $\ddcc = 5.0$ as a function of the temperature. The black dotted line corresponds to the parametrization defined by \cref{eq:DeltaTcSLDAE} valid close to the critical temperature.
}
\label{fig:phae-diagram}
\end{figure*}

As a first application we provide the phase diagram produced 
by the SLDAE functional. Namely, the value of the pairing gap
within the $(T/T_\rmF, \ddcc)$ space (where $T_\rmF = \varepsilon_\rmF$ is the \name{Fermi} temperature), has been shown in \cref{fig:phae-diagram}. 
\update{In \gls{BCS} theory, we can show that the ratio of the pairing gap function with the critical temperature is the universal number $\Delta(T = 0)/T_c^{\textrm{BCS}} = \pi/\e^{\gamma} \simeq 1.764$ where $\gamma$ is the \name{Euler} constant.}
The calculations provide a superfluid critical temperature [white solid line] above the \gls{BCS} theory result
[white dashed line]. \update{The first one is defined as temperature at which $\Delta$ reaches zero, while the $T_c^{\textrm{BCS}}$ we obtained through the formula but using the value of the self-consistent paring gap obtained numerically from our SLDAE functional.} 
The result suggests that the physical quantities can be expressed as an expansion of the ratio $\Delta / \varepsilon_\rmF\ostar$. 
The {SLDAE} functional predictions follow the \gls{BCS}-like self-consistent equations in such a way that universal relationships in \gls{BCS} theory are fulfilled up to the first order in ${\Delta/\eF\ostar}$. 
It is the case, in particular, for the critical temperature. 
Thus, using our \gls{EFT} correspondence, \respadd{we expect that the critical temperature can be approximated as}
\begin{subequations}
\begin{align}\label{eq:TcSLDAE}
    \frac{T_c}{T_\rmF} \approx \frac{\e^\gamma}{\pi} \frac{\Delta(T = 0)}{\varepsilon_\rmF\ostar}
    + \order*{\frac{\Delta(T=0)}{\eF\ostar}}^2,
\end{align}
where the higher order terms are neglected in the expansion. \respadd{Indeed, we observe reasonable 
accuracy of this formula when compared to} numerical calculation for all values of the density-dependent coupling constant \respadd{[left panel of fig.~\ref{fig:phae-diagram}]}
One may also note that, close to the critical temperature, the pairing gap function 
is in a good agreement with the \gls{BCS} asymptotic universal relation along the whole range of $\ddcc$ [dotted black line \update{in the right panel of fig.~\ref{fig:phae-diagram}}], \ie
\begin{align}\label{eq:DeltaTcSLDAE}
    \frac{\Delta(T \sim T_c)}{\eF\ostar} \sim \frac{T_c}{T_\rmF} \sqrt{\frac{8\pi^2}{7\zeta(3)}\qparen{1 - \frac{T}{T_c}}},
\end{align}
where $\zeta(s)$ is the \name{Riemann} zeta function.
\end{subequations}

Despite a systematic improvement of the physical quantities obtained through the expansion in $\Delta/\eF\ostar$, it is well known that \gls{SLDA}-like functionals overestimate the critical temperature observed in experiments and obtained in \abinitio calculations (see table VI and VII of \cite{Boettcher2014} for an extensive overview).
The interpretation of this is that we neglect \gls{BMF} effects, \ie we assume that the pairing gap function is proportional to the anomalous density only. Thus, as mentioned, using the \gls{EFT} formulation of the \gls{SLDA}, further investigation could be made to go beyond finite-temperature \gls{HFB} approximation by the use of perturbative techniques. Another strategy could be also to allow a temperature dependence of the density-dependent coupling constants appearing in the functional, in such a way that the critical temperature is well reproduced.
All of these considerations are out of the scope of this work if we restrict our studies to systems at low-temperature for which we expect good reproduction of the physical properties.
In the next section, as an example of applications of our developments, we discuss the static properties of superfluid vortices.

\subsection{Single superfluid vortex state properties}
Recent observations of gravitational waves during the merger of neutron stars \cite{Abbott2017} have led to a resurgence of interest during the last years on superfluid vortices. 
According to numerical simulations, during the fusion, one observes that a shear interface develops and involves \name{Kelvin} -- \name{Helmholtz} instabilities forming a series of vortices \cite{Obergaulinger2010,Giacomazzo2011,Kiuchi2014,Kiuchi2015} whose dynamic processes are still poorly understood. 
Besides, superfluid vortices are at the core of the superfluid property of the matter in general. 
To cite some selected examples of emergent phenomenon involving quantum vortices, we can mention the \name{Abrikosov} lattices \cite{Huebener2019,Kopycinski2021}, the \name{Onsager} -- \name{Kolmogorov} energy cascade in quantum turbulence \cite{Polanco2020,Skaugen2017,Reeves2013,Kobuszewski2020}, the vortex reconnection process \cite{Tylutki2021}, the pulsar glitches \cite{Pecak2021,Haskell2015}, \etc
It turns out that experimental realizations of such systems in dilute ultracold \name{Fermi} systems are currently investigated \cite{Kwon2021}. 
One of the observations is the dissipation occurring during the collision of two vortices, even in purely superfluid state at low-temperature.
This offers new opportunities to compare the theoretical understanding of the underlying processes with the observations. 
The development presented in this article can link easily with experiments in which the interaction can be fine-tuned continuously. 
It is therefore appropriate and timely to study the properties of such topological defects in superfluid systems.
Thus, we propose below to provide our predictive results on the static structure of superfluid vortices which constitute the first step towards fully large-scale dynamical simulations from \gls{BCS} to unitary regimes to know if \gls{SLDA}-like functional can quantify properly dissipation processes occurring in many-vortex systems.

In order to study the structural properties of such systems, we have first generalized the functional to the non-uniform case (see \cref{app:numerical-implementation} for details). We also have added the current density terms enforcing the Galilean invariance of the SLDAE functional (see discussion in section~\ref{sec:local-DFT}).
Then, we had considered imprinted superfluid vortex, as usually done in the calculations made with the \WSLDA, at the center of a tube-like trapping potential, periodic along the vortex line axis $r = 0$.
Below we present our results on the typical scales of the systems of interest which are summarized in \cref{tab:sv-scale}.

\begin{table*}[]
\caption{
\label{tab:sv-scale}
Properties of superfluid vortices at $T = 0.05 \,T_\rmF$ for selected value of the $s$-wave scattering length obtained using the SLDAE functional. The length scales of the vortex state (density at the center of the vortex core $n_v$ according to the bulk density $n_0 = \kF^3/3\pi^2$, the coherence length $l_c$ obtained with \cref{eq:lc-def}, and the vortex core radius $r_v$) are given by the first block, and the energy scales (the pairing gap in the bulk $\Delta_0$, the mini-gap energy $E_\mathrm{m.g.}$, and the critical temperature $T_c$) are provided by the second block.
{The error bars for the vortex core radius $r_v$ are due to lattice spacing uncertainty.}}
\begin{tabularx}{\textwidth}{@{\extracolsep{\fill}}llrrrrrrrrrr@{}}
\toprule
$\lambda$ &
   &
  $1.00$ &
  $1.50$ &
  $2.00$ &
  $2.50$ &
  $3.33$ &
  $5.00$ &
  $10.0$ &
  $20.0$ &
  $50.0$ &
  $\infty$ \\ \midrule
$n_v$           & $[n_0]$                    & 0.963 & 0.849 & 0.718 & 0.623 & 0.524 & 0.427 & 0.337 & 0.296 & 0.274 & 0.262 
 \\
$l_c$           & $[k_\mathrm{F}^{-1}]$      & 5.519 & 2.874 & 2.241 & 1.950 & 1.706 & 1.503 & 1.342 & 1.282 & 1.253 & 1.238 
 \\
$r_v$           & $[k_\mathrm{F}^{-1}]$      & 9.4(1) & 3.7(1) & 2.5(1) & 2.1(1) & 1.7(1) & 1.4(1) & 1.2(1) & 1.1(1) & 1.0(1) & 1.0(1) \\ \midrule
$|\Delta_0|$    & $[\varepsilon_\mathrm{F}]$ & 0.108 & 0.201 & 0.251 & 0.283 & 0.317 & 0.351 & 0.388 & 0.408 & 0.422 & 0.431 
 \\
$E_\text{m.g.}$ & $[\varepsilon_\mathrm{F}]$ & 0.009 & 0.018 & 0.034 & 0.048 & 0.066 & 0.087 & 0.112 & 0.127 & 0.137 & 0.144 
 \\
$T_c$           & $[T_\mathrm{F}]$           & 0.085   & 0.137  & 0.173  & 0.199  & 0.227  & 0.259  & 0.291  & 0.304  & 0.309  & 0.311  \\ \bottomrule
\end{tabularx}
\end{table*}

\subsubsection{Energy scales of superfluid vortex}
\begin{description}
    \item[The temperature] is a measure of the typical thermal excitation energy of the systems.
     The temperature of the quasi-particle states is set to $T = 0.05\,T_\rmF$ (typical temperature accessible in experiments) such that superfluid component of the gas vanish for $\ddcc \lesssim 1$. In the following, the temperature dependence of quantities will be implicit. 
    \item[The pairing gap] is the minimal energy of the quasi-particles in absence of topological defects. 
    The pairing gap at the center of the vortex becomes zero. The finite temperature effect does not affect this property as showed on \cref{fig:density-profile}(b) where $\abs{\Delta(r=0)/\Delta_0} \ll 1$ with $\Delta_0$ denoting the bulk pairing gap function. 
    \item[The mini-gap energy] is the typical energy scale of \name{Andreev} states, \ie the energy carried by the vortex core structure \cite{Andreev1964,Gennes1999,Sauls2018,Fisher2014,Eltsov2014,Pecak2021,Silaev2014}.
    Localized states exist in the vortex core due to the \name{Andreev} reflections with energies below the gap energy $E_\mathrm{m.g.} < E < \abs{\Delta_0}$. Using our \gls{EFT} correspondence with the \gls{BCS} theory, we can define the mini-gap energy as follows:
    \begin{align}\label{eq:Emg-def}
        \frac{E_\mathrm{m.g.}}{\eF} = \frac{1}{2} \abs*{\frac{{\Delta_0}}{\eF\ostar}}^2.
    \end{align}
    The numerical results extracted from our simulations at various values of the density-dependent coupling constant [blue circle] for the mini-gap energy displayed in \cref{fig:Emg}(a) are in good agreement with this definition [green solid line].
    {We observe discrepancy close to unitarity that we interpreted as due to the fact that (i) we neglected higher order correction in $\abs{\Delta_0}/\eF\ostar$ in  \cref{eq:Emg-def}, (ii) to the \gls{EFT} truncation used to design the SLDAE, and (iii) to the intrinsic errors induced by the regularization scheme.}
\end{description}

\subsubsection{Length scales of superfluid vortex}
\begin{figure}
\centering
\includegraphics{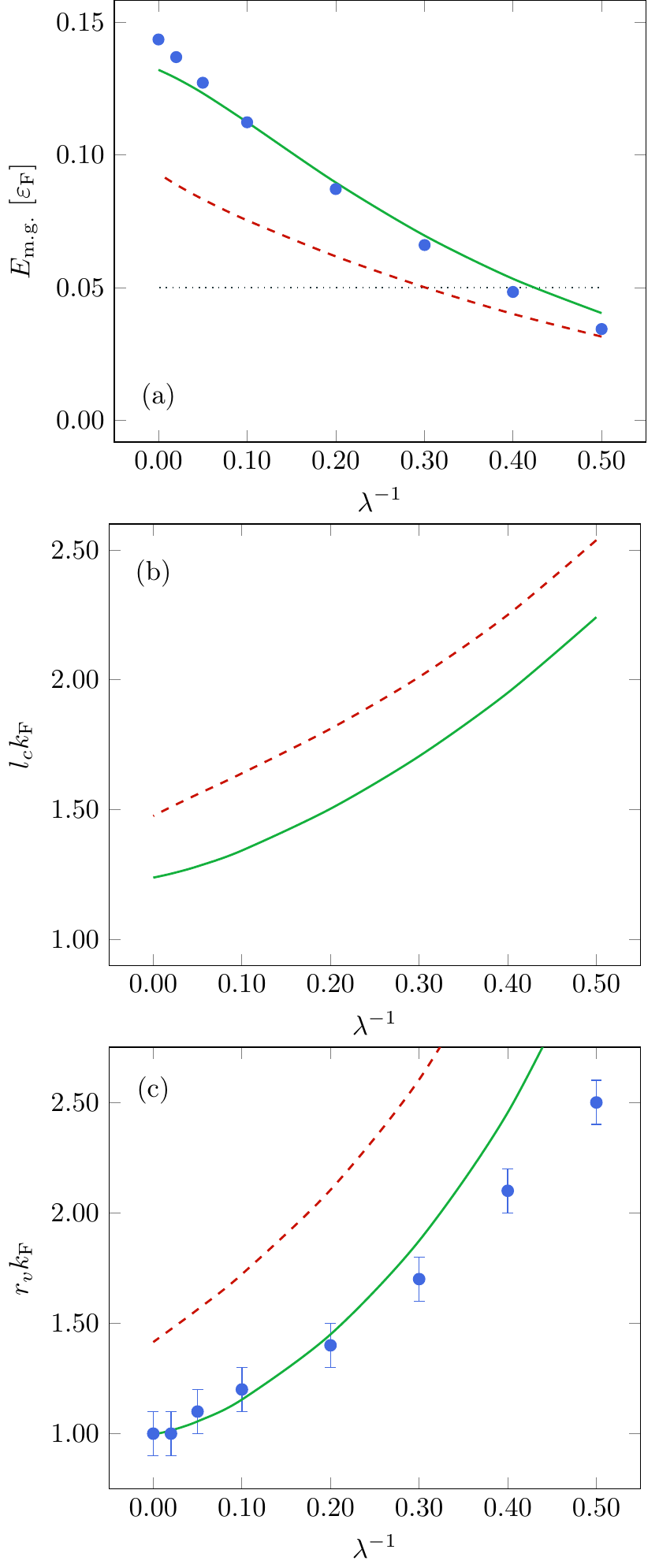}
\caption{
Mini-gap energy (a), coherence length (b), and vortex core radius (c) for a single vortex obtained using the SLDAE functional at temperature $T = 0.05\,T_\rmF$ as a function of $\ddcc\mo$ [blue filled circle]. 
{The error bars for the vortex core radius $r_v$ are due to lattice spacing uncertainty.}
For comparison, the red dashed line correspond to the \gls{BCS} result and the green solid line correspond to the associated one using our \gls{EFT} correspondence defined by \cref{eq:Emg-def,eq:lc-def,eq:rv-def} respectively.
We observe discrepancy according to the \gls{EFT} correspondence at $\ddcc\mo \gtrsim 0.3$ for the vortex core radius since the temperature [represented by the dotted black line in (a)] is above or close to the mini-gap energy. 
}
\label{fig:Emg}
\end{figure}

\begin{figure}
\centering
\includegraphics{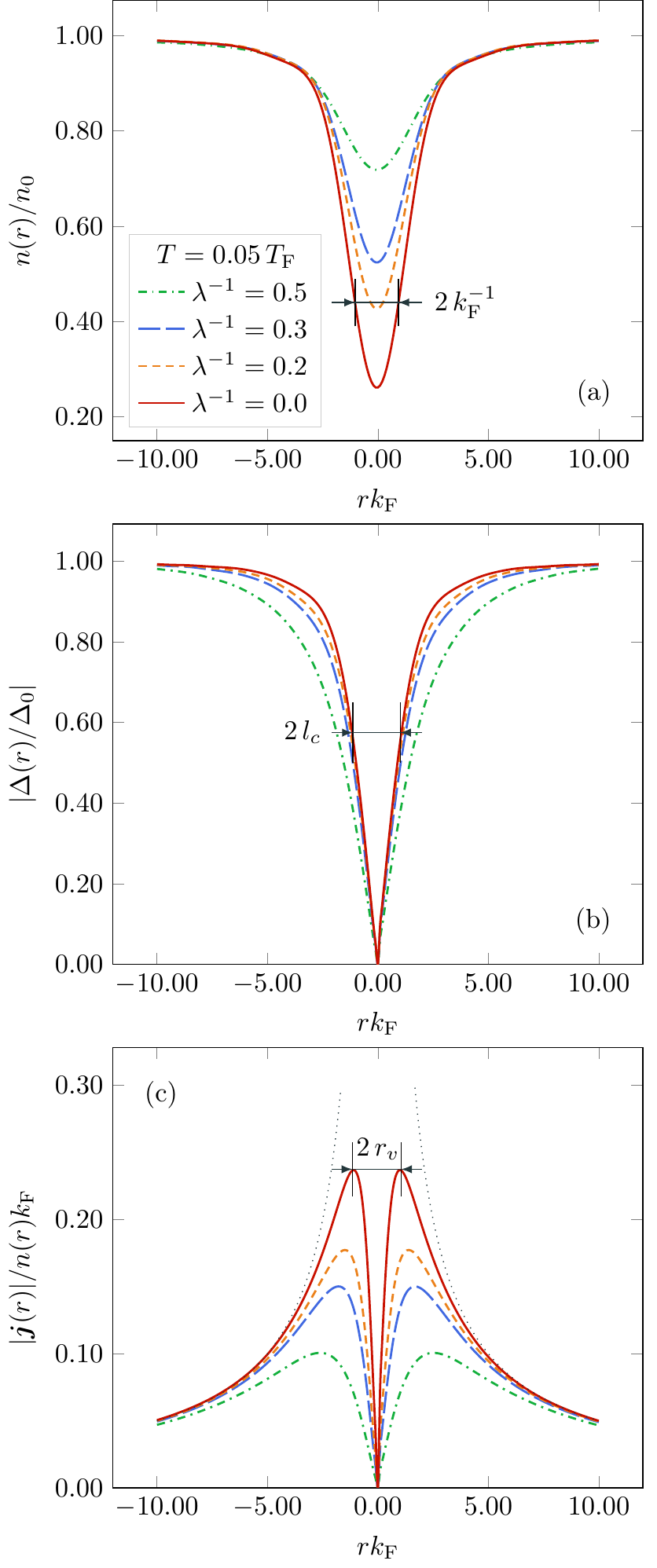}
\caption{
Density profiles of a single vortex at $T = 0.05\,T_F$ for several value of the density coupling constant $\ddcc$: (a) normal density, (b) pairing gap function, and (c) current density. The arrows show length scales of the system at unitarity: the inverse \name{Fermi} momentum, the coherence length, and the vortex core radius.
{For reference, the superfluid velocity $v_s = 1/2r$ is displayed [dotted black line].}
}
\label{fig:density-profile}
\end{figure}

\begin{description}
    \item[The \name{Fermi} momentum]
    is the characteristic length associated to the variation of the density [see \cref{fig:density-profile}(a)].
    We set our calculations in such a way that, for various density-dependent coupling constant $\ddcc$, the \name{Fermi} momentum $k_\rmF$ is obtained from the bulk density $n_0 = \kF^3/3\pi^2$. 
    As shown in \cref{fig:density-profile}(a) and \cref{tab:sv-scale}, the density at the center of the vortex line, $n_v$, reaches the bulk value in the weak coupling regime. 
    In particular, below $\ddcc \lesssim 1$, that is to say close to the critical temperature, the vortex vanish identically. 
    \item[The coherence length]
    is the characteristic length of the pairing gap variation [see figure \cref{fig:density-profile}(b)].
    As we have proved above, our \gls{EFT} correspondence provides a good parametrization of the physical quantities. 
    Following the same idea, we define the coherence length of the system as
    \begin{align}\label{eq:lc-def}
        l_c = \frac{2\eF\ostar}{\pi \kF\abs{\Delta_0}}
    \end{align}
    and compared to the \gls{BCS} theory result in \cref{fig:Emg}(b).  Note that the difference between our \gls{EFT} correspondence and the \gls{BCS} results are due to the effective mass term. Strictly speaking, considering $m\ostar > m$ leads to a smaller \name{Cooper} pair size characterized by the coherence length.
    \item[The vortex core radius]
    is the characteristic length scale for the superflow variation [see \cref{fig:density-profile}(c)].
    Finally, we define the vortex core radius when the current density reaches its maximum, \ie ~ 
    {$\abs{\vbj(r_v)} \equiv \max\, \abs{\vbj(r)}$}.
    Guided by the so-called \name{Ginzburg} -- \name{Landau} theory of phase transitions \cite{Terhaar1965,Gorkov1959,Abrikosov1957,Chaikin2000,Landau2013,Landau1980}, we argue that the ratio between the coherence length with the radius of the vortex core (similar to the penetration depth of superconductor in case of superfluid vortex) $\kappa \equiv r_v / l_c$, known as the \name{Ginzburg} -- \name{Landau} parameter, is temperature-independent, \ie the system is scale-invariant at the phase transition. 
    For instance, $0 < \kappa < 1/\sqrt{2}$ corresponds to type I superconductors while $1/\sqrt{2} < \kappa$ corresponds to type II superconductors.
    It turns out the equation
    \begin{align}\label{eq:rv-def}
        \kappa \equiv \frac{r_v}{l_c} \simeq \frac{1}{4}\frac{T_\rmF}{T_c}
    \end{align}
    provides an accurate approximation of the numerical results [blue circle] when $E_\mathrm{m.g.} \gtrsim T$ as showed in \cref{fig:Emg}(c) [green solid line]. 
    Further careful investigations are envisioned to conclude about the validity of our approximation in a realistic range of low-temperature.
\end{description}

The main conclusion that we can extract from our calculations is that typically the BCS-type formulas work well, once the \gls{EFT} correspondace is applied, i.e. $\Delta/\eF\rightarrow\Delta/\eF\ostar$. The SLDAE functional aims to be accurate at quantitative level for all values of $\ddcc$ (at low temperatures), and provided above results are the first step toward its validation. 
In particular we demonstrate that in experimentally realizable BCS regime ($\ddcc\mo  \approx 0.3 $ and $T \approx 0.05\,T_\rmF$) we have
$E_\mathrm{m.g.} \approx T$, which means that the impact of thermal effects cannot be neglected when analyzing dynamics of the vortex.  
This implies that in this regime, additional dissipation processes due to a non-vanishing shear viscosity are at play \cite{Wlazlowski2013a} {due to thermal excitation of the \name{Andreev}'s states leading to an increasing of the vortex core radius as observed in \cref{fig:Emg}(c).} 
On the strong interacting side, thermal effects are negligible and we expect to be able to study quantitatively dynamical processes causing the observed dissipation in the future time-dependent extension of the functional (work in progress). 
    
\section{Conclusion}

In this work, we have developed a systematic extension of the \gls{SLDA} from \gls{EFT} perspective, valid from the \gls{BCS} regime to the unitarity implying the bare density-dependent coupling constant $\ddcc = \abs{a_sk_F}$ only. 
Starting from the quasi-particle properties extracted from 
\abinitio calculations and/or experiments at zero-temperature for the associated homogeneous dilute system, the general solution of the \gls{BCS} equations are given as an expansion in $\Delta/\eF\ostar$ allowing us to deduce the functional parameters entering into the local \gls{DFT}. 
The clear advantage of this strategy consists to make the fitting procedure of functional parameters unnecessary and rendering this approach applicable to a large range of systems under the same parametrization relying on selected physical quantities.
Also, relying on a \gls{EFT} picture, we have identified $\Delta/\eF\ostar$ as been the proper parameter of the \gls{EFT} expansion for \gls{SLDA}-like \gls{DFT}.

This success allowed one to use standard local density approximation, \ie admit a spatial dependence of the densities, to study non-uniform systems. 
We have then applied our SLDAE functional through the numerical resolution of local generalized \gls{BdG} equations. 
Calculations of static properties of superfluid quantum vortices for several values of the $s$-wave scattering length have been carried out in order to facilitate discussions, for instance, with groups doing experiments aiming to emphasize dissipation processes in many-vortex systems. 
These results should help to interpret future simulations on dynamical processes involving vortices. 
Moreover, the ongoing implementation of the SLDAE functional in time-dependent variant promises to be a powerful tool to study the collective behavior of superfluid systems (linear response, \name{Higgs} mode, quantum quenches, \etc).

For possible future developments, we argue that our formulation allows us to consider \gls{BMF} effects into the functional using the standard perturbation methods which will lead to generalized self-consistent \gls{BdG} equations.
We would like to finally address a message to the \abinitio and \gls{EFT} communities to improve further the approach presented in this article.
First of all, the developed method depends strongly on the quality of the density-dependent quasi-particle properties of the systems considered as, for example, the effective mass and the pairing gap functions. 
Consequently, high accuracy of such quantities is required to parametrize properly the functional. 
Then, the lack of an \gls{EFT} framework providing a proper description of many-body systems from first principles, \ie starting from the bare Hamiltonian only, limit opportunities for further developments in both directions. 
Despite the recent attempts in the developments towards an \abinitio formulation of the \gls{DFT} mentioned above, such formulations did not reach sufficient maturity in terms of predictive power to be reliable guides. {Recently, in paper~\cite{Medvedev2017} it was shown that an approach to the functional design based on ``constraint satisfaction'' is a necessary ingredient in the process of constructing a highly accurate energy density functionals. In this respect maintaining by new functionals close relation to underlying \abinitio approaches and analytical results is desirable.}

\section*{Acknowledgements}
  This work was supported by the Polish National Science Center (NCN) under Contracts No.~UMO-2017/26/E/ST3/00428 (AB,GW) and UMO-2017/27/B/ST2/02792 (PM). We also acknowledge Poznan Supercomputing and Networking Center (Poland) for providing us resources at the Eagle supercomputer (grant id: 518). Numerical implementation was supported by IDUB-
POB-FWEiTE-2 Project granted by the Warsaw University of Technology under the Program Excellence Initiative:
Research University (ID-UB).
\vfill
\appendix
\section{Approximation of \name{Legendre} functions \label{app:Legendre}}
    In this appendix, we provide accurate approximation of \cref{eq:BCS-Legendre} for $0<t<\abs{s}$ obtained form expansions of the \name{Legendre} functions $P_l(z)$ for $l = n/2$ with $n\in\bbN$. 
    First, we define $\bar{I}_l(t/\abs{s}) = I_l(s<0,t)/\abs{s}^l$ and up to $\order{t^6/s^6}$, \ie around the logarithmic singularity, we obtain the following:
\begin{widetext}
\begin{align*}
	\bar{I}_l(u)
	& = \qbrack{2 - \frac{lu^2}{2}(l-1) + \frac{lu^4}{32}(l-3)(l-2)(l-1)}
	    \bigg[{H_l + \f\ln{\frac{u}{2}}}\bigg]
    - \frac{u^2}{2}(1-l-l^2)
	+ \frac{u^4}{32} \qbrack{6 - 13l + \frac{3l^2}{2} + 5l^3 - \frac{3 l^4}{2}},
\end{align*}
\end{widetext}
where $H_{l}$ denotes the harmonic numbers: $H_{1/2} = 2-2\ln{2}$, $H_{3/2} = 8/3-2\ln{2}$, $H_{5/2} = 46/15-2\ln{2}$, \etc
    Using this approximation, we can solve \cref{eq:BCS-Legendre} with the expansions given in \cref{eq:bc-expand} leading to $\calB_{2n+1}(x) = \calC_{2n+1}(x) = 0$ and
\begin{allowdisplaybreaks}
\begin{align*}
	\begin{split}
		\calB_0(x) & = -1,
	\end{split}
	\\
	\begin{split}
		\calB_2(x) & = 
		- \frac{x}{4} + \frac{1}{8} + \frac{3\ln{2}}{4},
	\end{split}
	\\
	\begin{split}
		\calB_4(x)
		& = \frac{3x^2}{64}
		  + \frac{5x}{128}
		  - \frac{9x\ln{2}}{32}
		  + \frac{7}{512}
		  \\&
		  + \frac{27\ln^2{2}}{64}
		  - \frac{15\ln{2}}{128},
	\end{split}
	\\
	\begin{split}
		\calB_6(x)
		& = - \frac{7x^3}{384}
		    + \frac{21x^2\ln{2}}{128}
		    - \frac{37x^2}{1024}
		    - \frac{69 x}{4096}
		    \\&
		    - \frac{63x\ln^2{2}}{128}
		    + \frac{111x\ln{2}}{512} 
		    + \frac{209}{24576}
		    \\&
		    + \frac{63 \ln^3{2}}{128}
		    - \frac{333 \ln^2{2}}{1024}
		    + \frac{207 \ln{2}}{4096},
	\end{split}
	\\
	\begin{split}
		\calB_8(x)
		& = \frac{55x^4}{6144}
		  + \frac{43x^3}{1536}
		  - \frac{55 x^3 \ln{2}}{512}
		  + \frac{1395x^2}{65536}
		  \\&
		  + \frac{495x^2\ln^2{2}}{1024}
		  - \frac{129x^2\ln{2}}{512}
		  - \frac{2141x}{393216}
		  \\&
		  - \frac{495x\ln^3{2}}{512}
		  + \frac{387x\ln^2{2}}{512}
		  - \frac{4185x\ln{2}}{32768}
		  \\&
		  + \frac{8657}{3145728}
		  + \frac{1485\ln^4{2}}{2048}
		  - \frac{387\ln^3{2}}{512}
		  \\&
		  + \frac{12555\ln^2{2}}{65536}
		  + \frac{2141\ln{2}}{131072},
	\end{split}
	\\
	\begin{split}
		\calC_0(x) & = \frac{x}{4} + \frac{1}{2} - \frac{3\ln{2}}{4},
	\end{split}
	\\
	\begin{split}
		\calC_2(x)
		& = \frac{x^2}{32}
		  - \frac{3x}{16}\ln{2}
		  + \frac{1}{64}
		  + \frac{9\ln^2{2}}{32},
	\end{split}
	\\
	\begin{split}
		\calC_4(x)
		& = - \frac{x^3}{128}
		    + \frac{9x^2\ln{2}}{128}
		    - \frac{13x^2}{1024}
		    - \frac{9 x}{1024}
		    \\&
		    - \frac{27x \ln^2{2}}{128} 
		    + \frac{39x \ln{2}}{512} 
		    - \frac{15}{8192}
		    + \frac{27 \ln^3{2}}{128}
		    \\&
		    - \frac{117 \ln^2{2}}{1024}
		    + \frac{27 \ln{2}}{1024},
	\end{split}
	\\
	\begin{split}
		\calC_6(x)
		& = \frac{5x^4}{1536}
		  + \frac{29x^3}{3072}
		  - \frac{5x^3\ln{2}}{128}
		  + \frac{315x^2}{32768}
		  \\&
		  + \frac{45x^2\ln^2{2}}{256}
		  - \frac{87x^2\ln{2}}{1024}
		  + \frac{385x}{196608}
		  \\& 
	      - \frac{45x\ln^3{2}}{128}
		  + \frac{261x\ln^2{2}}{1024}
		  - \frac{945x\ln{2}}{16384}
		  \\& 
		  - \frac{701}{393216}
		  + \frac{135\ln^4{2}}{512}
		  - \frac{261\ln^3{2}}{1024}
		  \\& 
		  + \frac{2835\ln^2{2}}{32768}
		  - \frac{385\ln{2}}{65536},
	\end{split}
	\\
	\begin{split}
		\calC_8(x)
		& = - \frac{27x^5}{16384}
		    - \frac{111x^4}{16384}
		    + \frac{405x^4\ln{2}}{16384}
		    - \frac{1257x^3}{131072}
		    \\&
		    - \frac{1215x^3\ln^2{2}}{8192}
		    + \frac{333x^3\ln{2}}{4096}
		    - \frac{3695x^2}{1048576}
		    \\&
		    + \frac{3645x^2\ln^3{2}}{8192}
		    - \frac{2997x^2\ln^2{2}}{8192}
		    + \frac{11313x^2\ln{2}}{131072}
		    \\&
		    + \frac{5251x}{4194304}
		    - \frac{10935x\ln^4{2}}{16384}
		    + \frac{2997x\ln^3{2}}{4096}
		    \\&
		    - \frac{33939x\ln^2{2}}{131072}
		    + \frac{11085x\ln{2}}{524288}
		    - \frac{6223}{8388608}
		    \\&
		    + \frac{6561\ln^5{2}}{16384}
		    - \frac{8991\ln^4{2}}{16384}
		    + \frac{33939\ln^3{2}}{131072}
		    \\&
		    - \frac{33255\ln^2{2}}{1048576}
		    - \frac{15753\ln{2}}{4194304}.
	\end{split}
\end{align*}
\end{allowdisplaybreaks}


   


\section{Improved \gls{BdG} functional \label{app:improved-BdG}}
    Following the strategy developed in this work, we propose to introduce an improved version of the \gls{BdG} functional defined by \cref{eq:BdG-functional}.
    For this, we start with the general local functional form \eqref{eq:sldae-functional}:
\begin{align}\label{eq:iBdG-functional}
	\mathcal{E}
	& = \bar{A}_\ddcc \frac{\tau}{2}
	  + \frac{3}{5}\bar{B}_\ddcc  n \varepsilon_F
	  +  \frac{\bar{C}_\ddcc}{n^{1/3}} \abs{\nu}^2,
\end{align}
where the \gls{HFB}, \gls{SLDA}, and functional parameters,  denoted with a bar, are defined using the weak coupling limit, \ie \gls{MBPT} for dilute \name{Fermi} systems, of the associated parameter of the functional developed in the main text, \ie $\xi_\ddcc \to \bar{\xi}_\ddcc \sim 1 + \order{\ddcc}$, 
$\zeta_\ddcc \to \bar{\zeta}_\ddcc \sim 1+ \order{\ddcc}$, 
$\alpha_\ddcc = a_\ddcc \to \bar{\alpha}_\ddcc = \bar{a}_\ddcc \sim 1+ \order{\ddcc}$,
and $\eta_\ddcc \to \bar{\eta}_{\ddcc} = (8/\e^2)\exp(-\pi/2\ddcc)$.

    Considering the first order of \cref{eq:bc-expand}, the \gls{HFB} parameters are given by:
\begin{subequations}
	\begin{align}
		\frac{\bar{b}_\ddcc}{\bar{a}_\ddcc} & = -1 + \order{\bar{x}_\ddcc^2},   
		\\
		\frac{\bar{a}_\ddcc}{\bar{c}_\ddcc}
		& = - \frac{\pi}{8 \ddcc} - \frac{\ln \bar{\alpha}_\ddcc}{4} + \order{\bar{x}_\ddcc^2}  ,   
	\end{align}
	with $\bar{x}_\ddcc \equiv \bar{\eta}_\ddcc/\alpha_\ddcc$.
\end{subequations}
    Then the associated functional parameters are obtained using \cref{eq:functional-functions}.
    For instance, up to second order of \gls{MBPT} for dilute \name{Fermi} gas, we have:
\begin{subequations}
	\begin{align}
		\bar{\xi}_\ddcc 
		& = 1 + p_1 \ddcc + p_2 \ddcc^2 +\order{\ddcc^3}
		\\
		\bar{\alpha}_\ddcc 
		& = 1 + q_1 \ddcc + q_2\ddcc^2 +\order{\ddcc^3}
	\end{align}
with $p_1 = -10/9\pi$, $p_2 = 4(11-2\ln{2})/21\pi^2$, $q_1 = 0$, and $q_2 = 8(1-7\ln{2})/15\pi^2$.
\end{subequations}
    This leads using \cref{eq:functional-functions}, up to second order in $\ddcc$, to the following functional parameters in \gls{DR} + \gls{MS}:
\begin{subequations}
	\begin{align}
		\bar{B}_\ddcc 
		& = \qparen{p_1 - q_1}\ddcc 
		  + \qparen{p_2 - q_2}\ddcc^2 ,
		\\
		\frac{n^{1/3}}{\bar{C}_\ddcc}
		& = \frac{1}{4\pi a_s}
		\qbrack{1 +  q_1\ddcc
		+ \qparen{\frac{2}{\pi}q_1 + q_1^2 - q_2}\ddcc^2}.
	\end{align}
\end{subequations}
    The functional obtained above corresponds to the weak coupling regime, \ie the limit $\ddcc \to 0$, of the main SLDAE functional designed in this work.


\section{Regularization of the contact interaction in EFT \label{app:EFT-reg}}
In this appendix, we propose a derivation of \cref{eq:slda-regularization} in the standard \gls{EFT} framework. 
We first define the \emph{in-vacum} regularization of contact interaction leading to the \gls{LEC} of the bare interaction.
Then, we derive similar renormalization for density dependent contact interaction leading to the \emph{in-medium} scheme used in this work.
    We will first recall generalities on the scattering theory and the renormalization of the loop integrals in \gls{EFT}. Then, we present renormalization procedure by considering the \emph{in-medium} effects and the presence of the \name{Fermi} sea.
    Note that our discussion will differs from others aspects of the regularization close to the \name{Fermi} surface in the context of the \name{Landau} theory of \name{Fermi} liquid as discussed in \cite{Furnstahl2008,Fitzpatrick2015,Polchinski1999,Shankar1994}.

\subsection{Generalities on scattering theory}

    We start with the leading order of a general non-relativistic local Lagrangian for a fermion field $\psi_0$ (with mass $m = 1$), invariant under Galilean, parity and time-reversal transformation. 
    Schematically, the Lagrangian reads:
\begin{align}
	\calL_0 & = \psi\odag_0 [\I \partial_t - \op{e}_{k} ]\psi_0 
	          - \frac{1}{2} \psi\odag_0 \psi\odag_0 \,\op{V}_0\, \psi_0 \psi_0 ,
\end{align}
where $\op{e}_{k} = (-\I\vbnabla)^2/2 \sim k^2/2$ is the Galilean invariant derivative.
    This Lagrangian is associated to the low-momentum effective $s$-wave interaction given by $V_0(k,k\oprime) = \mel{k\oprime}{\op{V}_0}{k} = g$.
    To connect the coupling constant to the standard \gls{LEC} of the bare interaction, we introduce the (on-shell) $S$-matrix for the $s$-wave scattering process, the associated (on-shell) $T$-matrix, and the phase shift $\delta$ defined as follows:
\begin{equation*}
	S_0(k) \equiv 1 - \frac{\I k T_0(k)}{2\pi} \equiv \e^{2\I \delta(k)} 
	\to
	T_0(k) = \frac{4\pi}{\I k - k \cot{\delta_0(k)}} .
\end{equation*}
The low-momentum expansion of the phase shift is given by $k\cot \delta_0(k) = -1/a_s +\order{k^2}$ that defines the $s$-wave scattering length $a_s$ and connects it to the constant $g$ of the bare Lagrangian. 
    Considering all orders in momentum, the $T$-matrix verifies the (on-shell) \gls{LSE} derived as follows.
    We consider the scattering of two particles interacting through the contact interaction $\op{V}_0 = g\delta(\vb{r}-\vb{r^\prime})$. Due to the fact that we consider the contact interaction, local in time, the two particles have necessarily different spins because of the \name{Pauli} exclusion principle.
    The states $\ket*{\psi^{\pm}}$ are the incoming ($-$) and outgoing ($+$) particles states solutions of $[\op{H_0} + \op{V}_0] \ket{\psi^{\pm}_k} = e_k \ket*{\psi^{\pm}_k}$.
    The initial and final \gls{sp} states are denoted by $\ket{\phi_k}$, and they are plane waves solutions of the \name{Schrödinger} equation $\op{H}_0\ket{\phi_k} = \epsilon_k \ket{\phi_k}$.
Due to the energy-momentum conservation, we have $e_k \to \epsilon_k = k^2/2$.
    Formally, the $S$-matrix is defined as $S(k,k\oprime)= \mel{\phi_{k\oprime}}{\op{S}}{\phi_{k}} = \braket{\psi_{k\oprime}^{-}}{\psi_{k}^{+}}$.
    We can show that
\begin{align} \label{eq:free-states-scattering}
	\ket{\psi_{k}^{\pm}} = \ket{\phi_k} + \op{G}_0^{\pm}(\omega = \epsilon_k) \op{V}_0 \ket{\psi_{k}^{\pm}},
\end{align}
where we have defined the free \acrlong{GF} or resolvent operator as $\op{G}_0^{\pm}(\omega) = [\omega - \op{H}_0 \pm \I\theta]\mo$.
    Then, the $T$-matrix is defined formally as $T_0(k,k\oprime) = \mel{\phi_{k\oprime}}{\op{V}_0}{\psi_k^+}$.
    Therefore, by inserting a closure relation $\ketbra{\phi_q}{\phi_q}$ in \cref{eq:free-states-scattering}, we deduce that the $T$-matrix verifies the \gls{LSE} given by
\begin{align} \label{eq:free-LSE}
	T_0(k\oprime,k) 
	& =  V_0(k\oprime,k)                           
	    - \frac{1}{4\pi^2} \int q^2\vd{q} \frac{V_0(k\oprime,q) T_0(q,k)}{e_q - \epsilon_k - \I \theta},
\end{align}
where $\epsilon_k = k^2/2$ denote the energy of the scattered asymptotic outgoing particle at infinity, \ie a free particle or plane wave. 
    Diagrammatically, this equation\footnote{Due the Galilean invariance, we can consider the center of mass frame in the loop calculations, \ie $\vbK = \vb{0}$.}\footnote{Note that we have to multiply the last diagrams by a factor $\I/2$ accounting for the symmetry factor of the diagrams, and by a factor $2$ accounting for spin summation, \cf Feynman rules in \cite{Hammer2000} for instance.} can be written as displayed in \cref{fig:diagrams-T-matrix}(a) where the propagator $G_0^{\pm}(\vbq,\omega) = [\omega - e_q \pm \I\theta]\mo$ is represented by the solid lines. 
    
    For the contact interaction, $V_0(k\oprime,k)=g$, the \cref{eq:free-LSE} can be solved analytically, and the solution reads
\begin{align} \label{eq:T0-contact}
	T_0(k) 
	& =  \frac{1}{\dfrac{1}{g}+\Lambda(k^2)},
\end{align}    
     where the explicit form of the loop integral $\Lambda$ is given in next section.

\subsection{Regularization of loop integrals \label{subsec:reg-loop}}
    In the last result, we have introduced the divergent loop integral
\begin{subequations}
	\begin{align}
		\Lambda(q^2) 
		& = \frac{1}{4\pi^2} 
		    \int \frac{k^2 \vd{k} }{e_k - \epsilon_q - \I \theta}         
		 \nonumber\\
		&= \frac{\I |q|}{4\pi}  
		    + \frac{1}{2\pi^2} \pv\int k^2 \vd{k} \frac{1}{k^2 - q^2} ,
	\end{align}
 where we have used $[X \pm \I\theta]\mo = \pv (1/X) \mp \I \pi \delta(X)$ with $\calP$ denoting the \name{Cauchy} principal value.
    In order to regularize this loop integral $\Lambda(q^2)$, a standard method is to insert a momentum scale $k_c$ and a regulator function $f(k/k_c)$ such that the integral converges. 
    This regulator satisfies $f(\infty) = 0$ and $f(0) = 1$.
    This defines the momentum dependent loop integral \cite{Birse1998}
    \begin{align}
	    \Lambda_c(q^2) 
	    & = \frac{1}{2\pi^2} \int \frac{k^2 \vd{k} }{k^2 - q^2 - \I \theta} \times f(k/k_c) .
    \end{align}
\end{subequations}

    We use a sharp-spherical cutoff prescription, \ie $f(k/k_c) = \varTheta(k_c - k)$ is a \name{Heaviside} step function, and we obtain 
\begin{align}
    \pv \Lambda_c(q^2) &= \frac{k_c}{2\pi^2} 
    \qbrack{1 - \frac{q}{2k_c} \ln\frac{k_c+q}{k_c-q}}.
\end{align}
    Note that, the loop integral still diverges with $k_c$. 
Combining it with \cref{eq:T0-contact} and low momentum expansion of $T_0(k)$ we obtain
\begin{align}\label{eq:gc-vacum}
	\frac{1}{g_c} = \frac{1}{4\pi a_s} - \pv \Lambda_c(q^2=0),
\end{align}
where $g_c$ denotes regularized coupling constant, while $(4\pi a_s)\mo$ is LEC expressed in terms of the scattering length.
Here, we recognize the renormalization scheme introduced in \cref{eq:BdG-regularization}.

\subsection{Regularization of pairing coupling constant}
    We consider now the following (grand-canonical) Lagrangian:
\begin{align} \label{eq:inmedium-Lagrangian}
	\calL = \psi\odag [\I \partial_t - \op{\varepsilon}_k]\psi
	      - \frac{1}{2}\psi\odag\psi\odag\,\op{V}_\ddcc\, \psi\psi ,
\end{align}
where the \emph{in-medium} interaction  $\mel{k\oprime}{\op{V}_\ddcc}{k} = C_\ddcc/n^{1/3}$ can be identified to the \gls{LEC} and the \gls{sp} energies are $\varepsilon_k \sim \alpha_\ddcc e_k + b_\ddcc\eF{}$ in our case.
    We consider now the shifted \gls{sp} energies in the one-body part of the Lagrangian due to the fact that we have introduced the chemical potential (contained in the definition of the $b_\ddcc$ parameter) to fix the number of particles.
    In other words, that consists in measuring the energies relative to the Fermi sea.
    In that case, the  \gls{LSE} above must also be redefined using the change $\pm\I\theta \to -\mu\pm\I\theta$. 
It is a consequence of the \name{Pauli} blocking: the scattering occurs only above the \name{Fermi} sea since all the \gls{sp} states below \name{Fermi} surface are occupied.
Note that similar idea are encontered to derive the Cooperon or \gls{mb} \gls{LSE} at \gls{RPA} level \cite[chap. 9]{Salomon2012} or resummation of ladder \emph{in-medium} many-body diagrams \cite{Kaiser2011,Kaiser2013,Boulet2019a}.

    More precisely, we consider the scattering of two particles interacting through a density-dependent contact interaction $\op{V}_\ddcc = C_\ddcc/n^{1/3}\delta(\vb{r}-\vb{r^\prime})$ with a \gls{mb} system of fermions with \gls{sp} energies $\varepsilon_k$ associated to an Hamiltonian $\op{H}$.
    The incoming ($-$) and outgoing ($+$) particle states are now solutions of $[\op{H} + \op{V}_\ddcc - \mu] \ket*{{\psi}^{\pm}_k} =  \varepsilon_k \ket*{{\psi}^{\pm}_k}$. 
    The initial and final states are solutions of $[\op{H}-\mu] \ket{{\phi}_k} = \tilde{\epsilon}_k \ket{{\phi}_k}$ 
    and are assumed as plane waves before and after the scattering.
    By energy-momentum conservation, we have $\varepsilon_k \to \tilde{\epsilon}_k = (\epsilon_k -\mu) + \mu = e_k$, \ie with energies higher than the chemical potential of the many-body system\footnote{Otherwise, the scattered particle cannot be differentiated from the \gls{sp} of the \gls{mb} system.}.

\begin{figure*}
\centering
\includegraphics{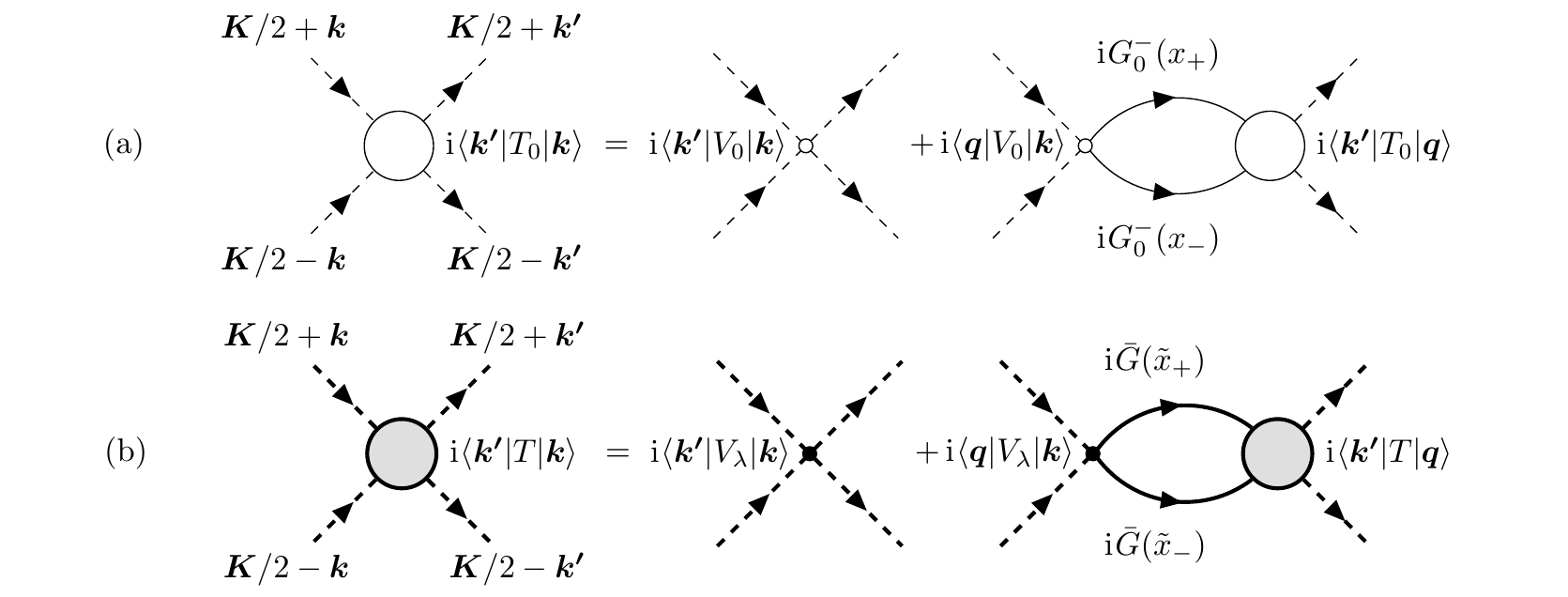}
\caption{Diagrammatic for the \emph{in-vacuum} (a) and \emph{in-medium} (b) $T$-matrix. 
For convenience, we have introduced the shorthand notations $x_\pm = (\vbq_\pm = \pm\vbk \pm \vbq, \epsilon_k \pm \omega)$ and $\tilde{x}_\pm = (\vbq_\pm, \tilde{\epsilon}_k \pm \omega)$.}
\label{fig:diagrams-T-matrix}
\end{figure*}

    We define the \gls{MBGF} as
\begin{align}
	G(\vbk,\omega) & = \frac{n_k}{\omega - \varepsilon_k - \I\theta} + \frac{1-n_k}{\omega - \varepsilon_k + \I\theta} 
	\nonumber 
	\\& \label{eq:mbGF}
	= \frac{1}{\omega - \varepsilon_k + \I \theta}
	+ 2\I\pi n_k \delta(\omega - \varepsilon_k)
\end{align}
decomposed into a free and a \emph{in-medium} components denoted respectively $\bar{G}$ and $\delta G$ and where $n_k$ denote occupation numbers of the particles in the medium.
    The \emph{in-medium} contribution, depending on occupation number, will not contribute to the \emph{in-medium} $T$-matrix, \ie we consider only the free propagation in the medium during the scattering process.
Therefore, as in the vacuum case, we get \cite{Rios2009}
	\begin{align}
		T(k,k\oprime) &= V_\ddcc(k,k\oprime) 
		\nonumber\\
		&+ \frac{1}{4\pi^2} \int q^2\vd{q} \bar{G}(\vbq,\omega = \tilde{\epsilon}_k)  V_\ddcc(k,q)T(q,k\oprime).
	\end{align}
    A diagrammatic representation of this equation is given in \cref{fig:diagrams-T-matrix}(b) where the thick arrowed solid lines correspond to the free component of the \gls{MBGF} defined by \cref{eq:mbGF} and the thick dashed arrowed lines to a free particle with a \gls{sp} energy above the \name{Fermi} sea.

    Finally, 
    we can formulate our \emph{in-medium} regularization procedure\footnote{This result leads to the identity $\pv\widetilde{\Lambda}_c \equiv  \pv\widetilde{\Lambda}_c(q^2 = 0) = \pv{\Lambda}_c(q^2 = -{2 b_\ddcc\eF}/\alpha_\ddcc)/\alpha_\ddcc$.
    In the vacuum, \ie at zero-density or equivalently in the limit $\ddcc \to 0$, we have $ - b_\ddcc\eF/\alpha_\ddcc \sim \kF^2 \to 0$ and $\alpha_\ddcc \to 1$, hence the continuity of the result since $\pv\widetilde{\Lambda}_c \to \pv\Lambda_c$.} 
\begin{subequations} \label{eq:renormalization-inmedium}
	\begin{align}
		\frac{n^{1/3}}{C_\ddcc^\mathrm{reg.}} 
		& = \widetilde{C}_\ddcc - \pv \widetilde{\Lambda}_c(q^2 = 0) ,
	\end{align}
	where:
	\begin{align} 
		\pv\widetilde{\Lambda}_c(q^2) 
		& = \frac{1}{4\pi^2} 
		    \pv\int 
		    \frac{k^2\vd{k}}{\varepsilon_k - \tilde{\epsilon}_q} ,
	\end{align}
\end{subequations}
by analogy to the result (\ref{eq:gc-vacum}).
Thus, we recover the results of \cref{eq:slda-regularization}.



\section{Numerical implementation of the SLDAE functional \label{app:numerical-implementation}}
    For simplicity of discussions, we keep in the main text the equations valid for uniform systems. 
    In case of non-uniform systems, \eg in an external potential, some modification of the self-consistent mean-field equations occurs. 
    We provide in this appendix details of our implementation of the general functional \cref{eq:sldae-functional} in the \WSLDA. 

\subsection{Non-uniform solutions}
    Considering the systems of interest trapped by an external static potential $V_\text{ext}(\vbr)$, the kinetic, potential, and pairing operators reads respectively \cite{Lipparini2003}:
\begin{subequations}
\begin{align}
    K(\vbr) &\equiv -\div \dvf{E}{\tau(\vbr)} \grad
    = - \frac{1}{2} \div A(\vbr) \grad,
    \label{eq:kinetic-operator}
    \\
    U(\vbr) &\equiv \dvf{E}{n(\vbr)},
    \\
    \Delta(\vbr) &\equiv - \dvf{E}{\nu\oast(\vbr)}
    = -\frac{C(\vbr)}{n(\vbr)^{1/3}} \nu(\vbr),
\end{align}
\end{subequations}
where we have introduced the shorthand notations $X(\vbr) \equiv X_{\ddcc(\vbr)}$ for the parameters which depend on the density dependent coupling constant $\ddcc(\vbr)$. 
    In particular, we anticipate regularization by replacing
\begin{align}\label{eq:Creg-fromCtilde}
    \frac{n(\vbr)^{1/3}}{C(\vbr)} \to
    \frac{n(\vbr)^{1/3}}{C^\text{reg.}(\vbr)}
    \equiv
    \widetilde{C}(\vbr) - \pv \widetilde{\Lambda}_c(\vbr).
\end{align}
where we note $\pv \widetilde{\Lambda}_c(\vbr) = \pv {\Lambda}_c(q(\vbr)^2) / A(\vbr)$ given explicitly below.
The effective coupling constant of the pairing part of the functional, defined by \cref{eq:Creg-fromCtilde}, can be obtained using 
\begin{align}
    \frac{n(\vbr)^{1/3}}{C^\text{reg.}(\vbr)}
    \equiv
    \frac{n(\vbr)^{1/3}}{C(\vbr)} - \frac{\pv {\Lambda}_c(q(\vbr)^2)}{A(\vbr)}.
\end{align}
    The mean-field potential is finally defined by
\begin{align}
    {U}(\vbr) &= \frac{1}{2}\dvp{A}{n}(\vbr) \tau(\vbr) +
    \bigg[B(\vbr) + \frac{3}{5}\dvp{B}{n}(\vbr)n(\vbr)\bigg] 
    \eF(\vbr)
    \nonumber \\&
    - \bigg[A(\vbr)  \dvp{\widetilde{C}}{n}(\vbr)
        -\dvp{\pv \Lambda_c}{n}(\vbr) \bigg] 
    \frac{\abs{\Delta(\vbr)}^2}{A(\vbr) }
    \nonumber \\&
    - \frac{1}{A(\vbr)}\bigg[\widetilde{C}(\vbr)\abs{\Delta(\vbr)}^2 +\Delta\oast(\vbr) \nu(\vbr)\bigg]\dvp{A}{n}(\vbr)
    \nonumber \\&
    + V_\text{ext}(\vbr),
    \label{eq:sc-potential}
\end{align}
where the local \name{Fermi} energy $\eF(\vbr)$ is defined thought the local \name{Fermi} momentum $\kF(\vbr)$ related to the normal density as $\kF(\vbr) = (3\pi^2 n(\vbr))^{1/3}$.

\subsection{Self-consistent regularization}
    In the text, we introduced a cutoff momentum $k_c$.
    However, for non-uniform system, the momentum of the \gls{qp} is no more a good quantum number. 
    Instead, we introduce a cutoff energy $E_c$ such that the summations of \cref{eq:bdg-densities} are performed only on \gls{sp} states labeled by $n$ such that $\abs{E_n} < E_c$.
    Guided by the \emph{in-medium} regularization scheme introduced in \cref{sec:regularization}, we choose the cutoff energy
\begin{align}
    E_c = A(\vbr) \frac{k_c(\vbr)^2}{2} + U(\vbr) - \mu,
\end{align}
that define a position dependent cutoff momentum $k_c(\vbr)$.
    We can now compute the cutoff integral as:
\begin{widetext}
\begin{align}
    A(\vbr)\pv \widetilde{\Lambda}_c(\vbr) 
    \equiv \pv {\Lambda}_c(q(\vbr)^2) 
    &= \frac{A(\vbr)}{4\pi^2}\pv\int_0^{k_c(\vbr)} \frac{k^2 \vd{k}}{A(\vbr) k^2/2 + U(\vbr) - \mu}
    \nonumber \\&
    =
    \begin{cases}
        \displaystyle\frac{k_c(\vbr)}{2\pi^2}\qbrack{1 - \frac{q(\vbr)}{2k_c(\vbr)} \ln\abs*{\frac{k_c(\vbr) + q(\vbr)}{k_c(\vbr) - q(\vbr)}}}
        & \text{if~}  \mu - U(\vbr)   > 0, \\[1.2em]
        \displaystyle\frac{k_c(\vbr)}{2\pi^2} \qbrack{1 + \frac{q(\vbr)}{k_c(\vbr)} \f\arctan{\frac{q(\vbr)}{k_c(\vbr)}} }
        & \text{if~}  \mu - U(\vbr) \leq 0,
    \end{cases}
\end{align}
\end{widetext}
where $q(\vbr)^2 \equiv {2 \abs{U(\vbr) - \mu}/A(\vbr)}$ is the positive pole of the cutoff integral. 
    Note that the term $A(\vbr)k^2/2$ appearing in the denominator of the cutoff integral integrand corresponds to the action of the kinetic operator of \cref{eq:kinetic-operator} in the dual reciprocal space, \ie $-\I\vbnabla = \vbk$ is the momentum of the \gls{qp} considered.




\bibliography{sldae-paper.bib}

\end{document}